\documentstyle[aaspp4]{article}

\begin{document}

\title{Relativistic and Newtonian core-shell models: analytical and 
numerical results}
 
\author{Werner M. Vieira and Patricio S. Letelier}

\affil{Departamento de Matem\'atica Aplicada, Instituto de 
Matem\'atica,
Estat\'{\i}stica e Computa\c{c}\~ao Cient\'{\i}fica, Universidade
 Estadual de Campinas, CP 6065,\\ 13081-970 Campinas, SP, Brazil}

\authoremail{vieira@ime.unicamp.br, letelier@ime.unicamp.br}


\begin{abstract}

We make a detailed analysis of the exact relativistic core-shell
 models recently proposed to describe a black hole or neutron star
 surrounded by an axially symmetric, hollow halo of matter, and 
in a seminal sense also galaxies since there are massive shell-like
 structures -- as for example rings and shells -- surrounding many
 of them and also evidences for many galactic nuclei hiding black
 holes. We discuss the unicity of the models in relation to their 
analyticity at the black hole horizon and to the full elimination 
of axial (conical) singularities. We also consider Newtonian and 
linearized core-shell models, on their own to account for dust shells
 and rings around galaxies and supernova and star remnants around
 their centers, and also as limiting cases of the corresponding
 relativistic models to gain physical insight.

Secondly, these models are generic enough to numerically study the
 role played by the presence/lack of {\it{discrete}} reflection
 symmetries about planes, i.\ e.\ the presence/lack of equatorial 
planes, in the chaotic behavior of the orbits. This is to be 
contrasted with the almost universal acceptance of reflection
 symmetries as default assumptions in galactic modeling. We
also compare the related effects if we change a true central
 black hole by a Newtonian central mass. Our main numerical 
findings are:\\
1- The {\it{breakdown}} of the reflection symmetry about the
 equatorial plane in both Newtonian and relativistic core-shell 
models does i) {\it{enhance}} in a significant way the chaoticity 
of orbits in
 reflection symmetric {\it{oblate}} shell models and ii) 
{\it{inhibit}} significantly also the occurrence of chaos in
 reflection symmetric {\it{prolate}} shell models. In particular,
 in the prolate case the lack of the reflection symmetry provides
 the phase space with a robust family of regular orbits that is
 otherwise not found at higher energies.\\
2- The relative extents of the chaotic regions in the relativistic
 cases (i.\ e.\ with a true central black hole) are significantly 
larger than in the corresponding Newtonian ones (which have just
 a $-1/r$ central potential).\\

\end{abstract}


\section{Introduction}

We make a more detailed study, both analytical and numerical,
 of some {\it{exact relativistic}} solutions we recently proposed
 to describe static, axially symmetric massive core-shell
 systems (Vieira and Letelier 1996a, 1997, hereafter VL1 and
 VL2, respectively). We add in this way to the efforts of
 modeling many situations of interest in astrophysics
 involving massive shell-like structures around centers, as
for example black holes or neutron stars surrounded by massive 
shell and ring remnants. A nice illustration of this possibility
 is offered by the famous Supernova 1987A plus its physical
 rings, see e.\ g.\ Panagia et al.\ (1996), Meyer (1997) and 
Chevalier (1997). At least in a seminal sense, the model could 
also describe galaxies since there are many of them exhibiting 
massive rings (Sackett and Sparke 1990, Arnaboldi et al.\ 1993,
 Reshetnikov and Sotnikova 1997) and shells (Malin and Carter 
1983, Quinn 1984, Dupraz and Combes 1987, Barnes and Hernquist
 1992), while many others possibly have galactic nuclei hiding
 black holes (see Kormendy and Richstone 1995 for a review).

Specifically, we implement here a monopolar core (a black hole
 in the relativistic case) plus an exterior shell of dipoles,
 quadrupoles and octopoles. Obviously, these multipoles are
 shell-like Legendre expansions i.\ e.\ their corresponding 
terms increase with the distance in the intermediate vacuum
between the core and the shell. Beyond its own applicability, 
the model also points to the possibility of a realistic 
description of any axially symmetric relativistic core-shell
 configuration whose approximation as a static system should
 be valid at some useful time scale, via a systematic multipolar
 Legendre expansion. In fact, a further step in this program
 was lastly started in Letelier and Vieira (1997) by giving
 stationarity (rotation) to the relativistic case of a monopolar
 core plus a purely dipolar shell. This is an important 
improvement since real celestial objects do rotate. The present
 analysis will be extended to rotating core-shell models in a
 forthcoming contribution.

Additionally, we consider the Newtonian counterpart of the
 relativistic model above, which was only sketched in the 
previous works, and show that they could describe core-shell 
systems ``per se'' interesting in astronomy. So, we can have
 massive circumstellar dust shells around certain types of 
stars as byproducts of their death, see e.\ g.\ Barlow et al.\ 
(1994) for an example of (irregular) circumstelar shells around 
luminous blue variables and a review by Groenewegen et al.\ 
(1998) about dust shells around carbon Mira variables. Another
 potential application for core-shell models is to the more 
speculative possibility of hollow galactic halos of dark matter
 made of neutrinos recently considered by Ralston and Smith 
(1991), Madsen (1991) and Barnes (1993). Moreover, we will see
 that the study of the relativistic core-shell model in parallel
 to its Newtonian counterpart is useful also to clarify the
physical content of the former one.

After Poincar\'e (1957) and the KAM theory (after Kolmogorov
 1954, Arnol'd 1963a, 1963b and Moser 1967)
 it became well established that non-integrability and hence 
chaos is a general rather than exceptional manifestation in
 the context of dynamical systems (modeling or not physical
 situations) (see Berry 1977). Given this ubiquitous fact,
 an important issue in astronomical modeling is to study in 
which extent in phase space chaoticity rises in models that
 are relevant to describe real systems and what are its 
consequences. For example, Binney (1982a) discusses the 
difficulties of constructing stationary self-consistent models
 when a significant fraction of orbits are irregular since 
they may not obey neither Vlasov's nor Jean's equations. Then, 
it is remarkable that a wide class of fully integrable potentials,
 the so called St\"{a}ckel potentials (Lynden-Bell 1962, de
 Zeeuw 1985 and de Zeeuw et al.\ 1986), are feasible starting 
points to describe, by themselves or by adding perturbations,
 real disc, elliptical or even triaxial galaxies. For more 
realistic tridimensional models there are evidences for 
rounder and smoother mass distributions generating only 
relatively small fractions of chaotic orbits (Schwarzschild 
1979, Binney and Tremaine 1987 and Evans et al. 1997), while
 flattening and/or sharpening of the mass distributions, for
 example through increasing triaxiality and/or putting cusps
and central masses to mimic black holes, tends to increase 
the chaoticity and force us to take it into account (Gerhard 
and Binney 1985, Schwarzschild 1993, Merrit and Fridman 1996,
 Norman et al. 1996, Merrit 1997 and Valluri and Merrit 1998).
 On the other hand, the emergence of chaos in two-dimensional 
models has more loose correlations with morphological aspects, 
see for example Binney and Spergel (1982), Richstone (1982),
 Binney (1982b), Gerhard (1985) and Sridhar and Touma (1997).

We address in this work also a numerical study about the 
chaoticity of orbits trapped in the bound gravitational 
zones between the core and the external massive shell.
 Beyond its own applicability as seen above, we take advantage 
of models that are generic within the class of axially
 symmetric core-shell distributions and moreover offer
 exact relativistic and Newtonian counterparts to be compared,
 in order to achieve some understanding about chaoticity
 related to two aspects: firstly, to the role of reflection
 symmetries shared by almost all models in astronomy and
 astrophysics, and secondly to the consequences of treating 
exact central black holes as just Newtonian central masses.

In the first part of this article (section \ref{analitico})
 we present analytical results concerning some properties
 like unicity and analyticity of the relativistic solutions
 themselves, in connection with both the linearized core-shell
 solutions and the corresponding Newtonian models taken as
 limiting cases.

In section \ref{sol} we enlarge the relativistic solution
 presented in VL1 for a monopolar core plus a shell of 
quadrupoles and octopoles in order to include a dipolar 
shell component. This complete solution was only outlined 
in VL2. In this later reference it was emphasized that the 
case of a core plus purely dipolar shell is physically
 non-trivial already in the Newtonian gravity, moreover, 
it was shown that the Newtonian dipolar case is integrable, 
in contrast with its chaotic relativistic counterpart, which 
allowed us to characterize the chaotic behavior in this case
 as an intrinsic general relativistic effect. Here, we take 
advantage of this striking difference of the dipolar case in
 order to reinforce the numerical conclusions presented in
 section \ref{caos} concerning the non-trivial dynamical 
role played by the presence/absence of the {\it{discrete}}
 reflection symmetry about the equatorial plane on the
 chaoticity of the orbits. 
 
In section \ref{reg} we proceed to a full elimination of
 axial (conical) singularities (hereafter CSs) from the
 relativistic model, which was only partially accomplished 
in VL1. In fact, CSs are not globally removable from neither
 static nor stationary many-body relativistic solutions
 since they are  self-consistently demanded by Einstein's
 equations to strut the otherwise unstable configuration
 against its own gravity (see Robertson and Noonan 1968 and 
Letelier and Oliveira 1998 for a recent review). So, in the
 present case the most we can do is really move all them
 outside the shell, thus obtaining a true vacuum in the
 intermediate space between the core and the shell. We
also discuss in this section the conditions for the
 Kruskal-type analyticity of the solutions at the  horizon of
 the central black hole and use both aspects to find the 
conditions under which the unicity of the models is assured.

In section \ref{RW} we present the linearization of the 
relativistic model in the multipole strengths via the so
 called Regge-Wheeler (RW) formalism. In particular, we 
see that our model exemplifies the fact that CSs survive
 to the linearization process, their presence in the
 intermediate vacuum being in fact an obstruction to the
 application of the RW formalism. This does reinforce the
 interpretation of CSs as a kind of singular matter
 distribution necessary to the dynamical consistency of
 static and stationary relativistic models.
However being intrinsecally relativistic manifestations,
 an adiabatic treatment of CSs in the Newtonian limit allows
 for estimates concerning the emission rate of gravitational
 waves by two coalescing rotating black holes (Araujo et 
al.\ 1998).

In section \ref{New} we discuss Newtonian core-shell models,
 on their own as well as limiting cases of the corresponding
 relativistic ones. We put the H\'enon-Heiles-like structure
 of the relativistic models in the appropriate astronomical
 context and
stress the physical content of the various terms present in 
them, in particular we see that the apparently naive constant
 relativistic solution does in fact hide a relativistic
 homoeoid (see Chandrasekhar 1987 for Newtonian homoeoids).

The numerical part of this article is presented in section
\ref{caos} and deals with a study of the chaoticity of orbits
 trapped in the intermediate vacuum between the core and the
 shell. Specifically, we shall explore in this section the
 fact that, due to its generality, axially symmetric 
core-shell expansions are particularly suitable to study
 the dynamical role played by a discrete symmetry of
 the model --- namely the reflection symmetry about an
 equatorial plane, on the chaoticity of orbits. The 
reflection symmetry is present in the model if and only
 if the shell is of even-type, i.\ e.\ {\it{iff}}
 $2^{2n}$-poles, $n=0,1,2,3,...$, occur in the shell
 expansion, and it is broken if we add any odd, 
$2^{2n+1}$-poles to the expansion. The hypothesis of 
reflection symmetry is a widely spread assumption in
 astronomical modeling and its twofold justification 
lies firstly in that too many real celestial objects to
 be modeled, for example stars themselves, star clusters 
and galaxies seem in fact very symmetric with respect to
 a middle plane. Another reason for this symmetry
 assumption is that the resulting model gets strongly
 simplified in both analytical and numerical aspects.
 Of course, we do not expect the reflection symmetry
 be realized in nature exactly, so it is relevant to
 search for possible detectable dynamical effects arising
 from deviations of that symmetry.

A strong additional motivation for this numerical study is
 to compare from the point of view of the chaotic behavior
 what happens when the presence of true central black holes
 are simplified by reducing them to Newtonian central masses.
 This is a commom practice in current modeling, as for
 example in Gerhard and Binney (1985), Sridhar and Touma
 (1997) and Valluri and Merrit (1998).
 
Surprisingly enough, the numerical findings show marked
 effects on the chaoticity of the orbits in the intermediate
 vacuum between the core and the shell linked to the
 presence/lack of the reflection symmetry in the relativistic
 as well as Newtonian models. We also find strong quantitative
 differences in the chaoticity manifested in both relativistic
 and Newtonian cases.

Finally, we present the conclusions with some discussion
 and prospects in section \ref{discussion}.

\section{Analytical results: exact, linearized and
 Newtonian core-shell models}\label{analitico}

\subsection{The exact model}\label{sol}

We deal here with static, axially symmetric models, for
 which the Weyl coordinates $(t,\rho,z,\phi)$ are the 
starting point:
\begin{equation}
ds^2=e^{2\nu}dt^2-e^{2\gamma-2\nu}[dz^2+d\rho^2]
-e^{-2\nu}\rho^2d\phi^2,\label{hh1}
\end{equation}
where $\nu$ and $\gamma$ are functions of $\rho$ and
 $z$ only.
Except where units are explicitly required (particularly
 in section \ref{New}), we use non-dimensional variables:
 $s \leftrightarrow s/L$ (idem for $\rho$ and $z$), and
 $t \leftrightarrow ct/L$ where $c$ is the light velocity
 in vacuum and $L$ is some convenient unit of length.
 Einstein's equations in absence of matter (vacuum)
reduce in this case to the usual Laplace equation for $\nu$
\begin{equation}
\nu_{,\rho\rho}+\frac{1}{\rho}\nu_{,\rho}+\nu_{,zz}=0,
\label{hh2}
\end{equation}
and the quadrature
\begin{equation}
d\gamma=\rho[(\nu_{,\rho})^2-(\nu_{,z})^2]\;d\rho+
2\rho\nu_{,\rho}\nu_{,z}\;dz\label{hh3}
\end{equation}
for $\gamma$.

Before proceeding, we mention that the Weyl coordinates
 should be considered somewhat deceiving if we insist
 in naively transferring their image contents about
 mass configurations to the Newtonian common sense.
 For example, the spherical shape of the horizon of
 a black hole is compressed into a bar in Weyl's 
coordinates, the interior portion of the black hole
 space-time being wholly discarded from the portrait.
 Apart from the well known fact that the relativistic 
context is not cast in a straight relation with the 
Newtonian one in strong regimes, they are mathematically
 sound as will be clear later when we will compare
 both versions of the core-shell models.

We pass to the prolate spheroidal coordinates 
$(t,u,v,\phi)$,
which have a direct link with the ``spherical''
 ones $(t,R,\theta,\phi)$ ($R \leftrightarrow R/L$)
 we will use later,
\begin{eqnarray}
&&u=R-1=\frac{1}{2}[\sqrt{\rho^2+(z+1)^2}+\sqrt{\rho^2+
(z-1)^2}]
{\mbox{,\hspace{0.4cm}$u \geq 1,$}}\nonumber\\
&&v=cos\theta=\frac{1}{2}[\sqrt{\rho^2+(z+1)^2}-
\sqrt{\rho^2+(z-1)^2}]
{\mbox{,\hspace{0.4cm}$-1\leq v\leq 1$}},\label{hh4}
\end{eqnarray}
or, in terms of $\rho,z$,
\begin{eqnarray}
&&\rho=\sqrt{(u^2-1)(1-v^2)}=\sqrt{R(R-2)}sin\theta
{\mbox{,\hspace{0.4cm}$R\geq 2,$}}\nonumber\\
&&z=uv=(R-1)cos\theta.\label{hh5}
\end{eqnarray}
Eqs.\ (\ref{hh2}) and (\ref{hh3}) are written in
 terms of $u,v$ as
\begin{equation}
[(u^2-1)\nu_{,u}]_{,u}+[(1-v^2)\nu_{,v}]_{,v}=0,\label{hh5b}
\end{equation}
\begin{eqnarray}
&&\gamma_{,u}=\frac{1-v^2}{u^2-v^2}[u(u^2-1)(\nu_{,u})^2-
u(1-v^2)(\nu_{,v})^2-2v(u^2-1)\nu_{,u}\nu_{,v}],\nonumber\\
&&\gamma_{,v}=\frac{u^2-1}{u^2-v^2}[v(u^2-1)(\nu_{,u})^2-
v(1-v^2)(\nu_{,v})^2+2u(1-v^2)\nu_{,u}\nu_{,v}].\label{hh5a}
\end{eqnarray}
In the prolate spheroidal coordinates $u$, $v$,
 Laplace equation (\ref{hh5b}) can be separated
 and solved in terms 
of standard Legendre
polynomials $Q_{\ell},P_{\ell}$ (see for example 
Moon and Spencer 1988, and, for relativistic applications, 
Reina and Treves 1976):
\begin{equation}
\nu (u,v)=\sum_{0}^{\infty} [a_{\ell}Q_{\ell}(u)+
b_{\ell}P_{\ell}(u)][c_{\ell}
Q_{\ell}(v)+d_{\ell}P_{\ell}(v)].\label{hh6}
\end{equation}

The particular solution picked out from the general 
one above is determined by the matter distribution
whose model is wanted. 
We are interested here in monopolar
 core plus shell-type models,
 so we are guided by the Newtonian case 
(to be detailed in section \ref{New}) to
 the specific solution of the form (up to third order)
\begin{equation}
2\nu=2\nu_0-2\kappa Q_0(u)+2{\cal{D}}P_1(u)P_1(v)+
(4/3){\cal Q}P_2(u)P_2(v)+(4/5){\cal O}P_3(u)P_3(v).
\label{hh7}
\end{equation}
$\nu_0$ is an integration constant and $Q_0$ describes
 the monopolar core, which either reduces to a black
 hole if we put $\kappa =1$ and identify $2L$ with
 the Schwarzschild radius of the core (in which case 
$(t,R,\theta,\phi)$ above are Schwarzschild's coordinates),
 or can be switched off by simply putting $\kappa =0$.
 The remaining terms correspond to the multipoles
 originated from the exterior shell of matter: dipole
 ${\cal D}$, quadrupole ${\cal Q}$ and octopole
 ${\cal O}$ (note the opposite sign convention for
 dipoles in this definition with respect to that
 appearing in VL2). The nontrivial character of shell 
dipoles in both Newtonian and general relativistic
 theories of gravity has been anticipated in VL2 while
 shells made of quadrupoles plus octopoles were
 considered in VL1. If non-monopolar
 cores should be considered, we should add to 
(\ref{hh7}) terms of the form 
$Q_{\ell}(u)P_{\ell}(v)$, $\ell \geq 1$.

Much more sophisticated multipolar relativistic treatments
 for core, shell and core-shell models are considered
 by Thorne (1980), Zhang (1986) and Suen (1986a, 
1986b), respectively, in terms of higher order 
systematic expansions of the metric in de Donder 
coordinates. Our approach is more modest at this 
stage and inspired only on the zeroth order, Newtonian
 limit of the full relativistic situation.

The odd multipoles of $v$ in (\ref{hh7}) (${\cal D}$
 and ${\cal O}$ in the present case) break the
 reflection symmetry about the plane $z=0$, and
 we shall show in section \ref{reg} that we need
 of both ${\cal D}$ and ${\cal O}$ being simultaneously
 either present or absent if we want to rule out
 conical singularities of the intermediate vacuum
 between the core and the shell. We rewrite (\ref{hh7}) as
\begin{equation}
\begin{array}{l}
2\nu=\kappa \log(\frac{u-1}{u+1})+P(u,v),\\
P=2\nu_0+2{\cal{D}}uv+\frac{1}{3}{\cal{Q}}(3u^2-1)(3v^2-1)
+\frac{1}{5}{\cal{O}}uv(5u^2-3)(5v^2-3).\label{hh8}
\end{array}
\end{equation}
After integrating (\ref{hh5a}) we obtain the 
corresponding solution for $\gamma$:
\begin{equation}
\begin{array}{l}
2\gamma=\kappa^2\log(\frac{u^2-1}{u^2-v^2})+Q(u,v),\\
Q=2\gamma_0+\gamma_D+\gamma_Q+\gamma_O+\gamma_{DQ}+
\gamma_{DO}+\gamma_{QO},\\
\gamma_D=4\kappa {\cal{D}}v-{\cal{D}}^2[u^2(1-v^2)+v^2],\\
\gamma_Q=-4\kappa{\cal{Q}}u(1-v^2)\\
{\mbox{\hspace{1.0cm}}}+(1/2){\cal{Q}}^2[u^4(1-v^2)(1-9v^2)
-2u^2(1-v^2)(1-5v^2)-v^2(2-v^2)],\\
\gamma_O=-(2/5)\kappa {\cal{O}}v[5(3u^2-1)(1-v^2)-4]\\
{\mbox{\hspace{1.0cm}}}+(3/100){\cal{O}}^2[-25u^6(1-v^2)
(5v^2+2v-1)(5v^2-2v-1)\\
{\mbox{\hspace{1.0cm}}}+15u^4(1-v^2)(65v^4-40v^2+3)
-3u^2(1-v^2)(25v^2-3)(5v^2-3)\\
{\mbox{\hspace{1.0cm}}}-v^2(25v^4-45v^2+27)],\\
\gamma_{DQ}=-4{\cal{D}}{\cal{Q}}uv(u^2-1)(1-v^2),\\
\gamma_{DO}=(3/10){\cal{D}}{\cal{O}}[u^2(5u^2-6)(1-v^2)
(1-5v^2)+v^2(5v^2-6)],\\
\gamma_{QO}=-(6/5){\cal{Q}}{\cal{O}}uv(u^2-1)(1-v^2)
[(5u^2-1)(3v^2-1)+2(1-v^2)].\label{hh9}
\end{array}
\end{equation}
This is the complete solution, which was outlined
 in VL2. The additional integration constant
 $\gamma_0$ will be fixed in the next section
 in connection with the elimination of conical
 singularities from the intermediate vacuum,
 while the integration constant $\nu_0$ will
 prove to be necessary in assuring analyticity
 to the full core-shell solution at the horizon 
in the absence of conical singularities. The 
terms proportional to $\kappa$ in $\gamma_D$,
 $\gamma_Q$ and $\gamma_O$ represent nonlinear
 interactions between the black hole and 
the external shell. 

\subsection{Unicity and smoothness requirements}
\label{reg}

We shall be interested in two classes of singularities
 of the Weyl spacetime.
The first one are the strong singularities that are
 located in the points wherein 
the scalar polynomial invariants of the curvature tensor blow up.
 For a static
 axially symmetric spacetime solution to 
the vacuum field equations we have only two
non-vanishing invariants.
 They are (Carminati and McLenaghan
 1991) $w_1\equiv 
\frac 18C_{abcd}C^{abcd}$ and $w_2\equiv
-\frac 1{16}C_{ab}^{cd}C_{cd}^{ef}C_{ef}^{ab}$ 
where $C^{abcd}$ is the Weyl
trace-free tensor, which for vacuum solutions 
coincides with the Riemann
curvature tensor. After some algebraic manipulations, they
reduce to
\begin{equation}
\begin{array}{l}
w_1=2\kappa ^2[3\sigma (\nu_{,z}^2+\rho^2
\nu_{,\rho}^2\nu_{,z}^2
-\nu_{,\rho}/\rho-\rho\nu_{,\rho}
\sigma ) +\nu _{,\rho}^2(1+2\rho\gamma _{,\rho}+3\rho
\nu _{,\rho})\\
{\mbox{\hspace{1.0cm}}}+\rho^2(\nu_{,z}^6+\nu _{,\rho}^6)
 +\nu _{,\rho z}(6\nu_{,\rho}\nu_{,z}
+\nu_{,\rho z}-2(\nu_{,z}\gamma_{,\rho}+
\nu_{,\rho}\gamma _{,z}))\\
{\mbox{\hspace{1.0cm}}}+\nu _{,\rho\rho}(3
\gamma _{,\rho}(1-2\rho\nu _{,\rho}) /\rho-
\nu _{,zz}+2\nu _{,z}\gamma _{,z})],\\ 
w_2=3\kappa ^3\rho^{-2}(\nu _{,\rho}/\rho-\sigma )[\sigma
( 3\rho^3\nu _{,\rho}^2\nu _{,z}^2+\rho 
( 1-3\rho\nu _{,\rho}) \sigma +2\nu
_{,\rho})  \\
{\mbox{\hspace{1.0cm}}}-\nu _{,z}^2( 
\gamma _{,\rho}+3\nu _{,\rho}( 1-2\rho\nu _{,\rho}))
+\rho^3( \nu _{,z}^6+\nu _{,\rho}^6)  \\
{\mbox{\hspace{1.0cm}}}+r\nu _{,\rho z}( 2\nu _{,z}
( 3\nu _{,\rho}( 1-\rho\nu _{,\rho}) +\rho\nu
_{,z}^2) +\nu _{\rho z})  \\
{\mbox{\hspace{1.0cm}}} +\nu _{,\rho\rho}( 3
\gamma _{,\rho}( 1-2\rho\nu _{,\rho}) +4r^2\nu
_{,\rho}^3-\rho\nu _{, zz}) ] \label{hh9a}
\end{array}
\end{equation}
where $\sigma \equiv \nu _{,z}^2+\nu _{,r}^2$ an
 $\kappa \equiv \exp 2\left( \nu
 -\gamma \right) $. In the present case, these
 scalars are singular on the position of the attraction 
center and  will be also singular in some directions 
of the spatial infinity (the specific directions will
 depend on the signs of ${\cal{D}}$, ${\cal{Q}}$, etc.),
 in other words they are singular in the
position of the sources of the gravitational field.

There is another class of singularities that do not
 show up in a simple way
 in the curvature, the so called conical
 singularities (CSs) (Sokolov and Starobinskii 1977). These
singularities are in some sense like the 
distributions related to a
low dimensional Newtonian potential, e.g., for
 an infinite massive wire the potential is
  $\phi=2\lambda \ln \rho$,  we have that the Laplace
 equation, the analog to the curvature, 
gives us $\nabla^2\phi =0$ for $\rho\not = 0$ while for
 $\rho =0$ we need to use some global 
property, for example Gauss
 theorem, to  get $\nabla^2\phi =4\pi\lambda\delta(\rho )$.
To be more precise, let us consider a conical surface, 
 $z=\alpha \rho$,  embedded in the usual
 Euclidean three dimensional space $ds^2=d\rho^2+\rho^2d
\varphi^2+dz^2$, so that we have on the cone
$ds^2=(1+\alpha^2)d\rho^2 +\rho^2 d\varphi^2=d\bar\rho^2 +
\beta^2\bar\rho^2 d \bar\varphi^2$ with $\bar\rho=
\sqrt{1+\alpha^2}\rho$ and
$\beta^2=1/(1+\alpha^2)$. Note that the coordinate
 range of $\bar\rho$ as well as $\varphi$  are
 the usual ones. The ratio between the arc of the
 circumference and its radius
 is $2\beta \pi$ in this case. If we 
compute the Riemann tensor for this last
 metric we find that
 $R_{\bar\rho \varphi \bar\rho \varphi}=0$ for 
$\bar\rho\not =0$. By using the Gauss-Bonnet 
theorem in this case, 
we can pick up the curvature singularity as 
being of the form $R_{\bar\rho \varphi \bar\rho
 \varphi} \propto \delta(\bar\rho)$.

It is well known that CSs
 arise when we consider Weyl's solutions near the
 symmetry axis (Robertson and Noonan 1968 and Letelier
 and Oliveira 1998a). They are interpreted as a
 geometric consequence of the presence of some kind 
of ``strut'', necessary to consistently prevent any
 static non-spherically symmetric model from
 collapsing due to self-gravity. In this sense, the
 uniqueness of the core-type, one-body solution for
 the (static) spherically symmetric Einstein's equations
 (namely, the Schwarzschild one) should relate to the
 fact that it is impossible to attach struts in a 
perfectly round metric.
For the Weyl solutions, we can always consider 
a small disk
  centered in the symmetry axis ($t=t_0, z=z_0, 
\rho=\epsilon, 0\leq\varphi<2\pi$) and impose that
 the ratio between the circumference and 
 its radius equals $2\pi$. This is the condition
 to eliminate 
CSs from the intermediate vacuum
 (then attaching them to the infinity), which amounts
 to impose on the function $\gamma$ the well known 
conditions of elementary flatness
\begin{equation}
2\gamma\mid_{\rho=0,|z|>1}\equiv 2\gamma\mid_{u=|z|,
v=\pm 1}=0.\label{hh10}
\end{equation}
These two conditions fix the constant $\gamma_0$ and 
impose an additional constraint on the {\it{odd}}
 shell multipoles (${\cal D}$ and ${\cal O}$ here) 
in the presence of the black hole:
\begin{equation}
2\gamma_0-{\cal D}^2-\frac{1}{2}{\cal Q}^2-
\frac{21}{100}{\cal O}^2-\frac{3}{10}{\cal D}
{\cal O}=0,\label{hh11}
\end{equation}
\begin{equation}
\kappa [{\cal D}+\frac{2}{5}{\cal O}]=0.\label{hh11a}
\end{equation}
The condition (\ref{hh11}) fixing $\gamma_0$ is necessary
 in all cases to rule out the conical singularities.
 From (\ref{hh11a}) we see that the former is also a
 sufficient condition in two cases: i) the shell of
 dust is left alone by switching off the core ($\kappa = 0$),
 and ii) in the presence of the black hole ($\kappa=1$),
 the shell is made only of even-type multipoles (${\cal Q}$
 here). We also see that it is possible to eliminate 
conical singularities for a single or pure shell component
 only if it is of the even type. Then, it follows that
 the core-shell dipole solution presented in VL1, as
 well as those with ${\cal O}\neq 0$ presented in VL2,
 all have conical singularities since they do satisfy 
(\ref{hh11}) but {\it{not}} (\ref{hh11a}) . If we include
 the necessary condition (\ref{hh11}) in (\ref{hh9}), 
we have the following for $\gamma$:
\begin{equation}
\begin{array}{l}
2\gamma=\kappa^2\log(\frac{u^2-1}{u^2-v^2})+Q(u,v),\\
Q=\gamma_D+\gamma_Q+\gamma_O+\gamma_{DQ}+\gamma_{DO}+
\gamma_{QO},\\
\gamma_D=4\kappa {\cal{D}}v-{\cal{D}}^2(u^2-1)(1-v^2),\\
\gamma_Q=-4\kappa{\cal{Q}}u(1-v^2)-(1/2){\cal{Q}}^2
(u^2-1)(1-v^2)[u^2(9v^2-1)+1-v^2],\\
\gamma_O=-(2/5)\kappa {\cal{O}}v[5(3u^2-1)(1-v^2)-4]\\
{\mbox{\hspace{1.0cm}}}-(3/100){\cal{O}}^2(u^2-1)(1-v^2)
[(25u^4-20u^2+1)(25v^4-14v^2+1)\\
{\mbox{\hspace{1.0cm}}}+30u^2v^2(5v^2-1)+6(1-v^2)],\\
\gamma_{DQ}=-4{\cal{D}}{\cal{Q}}uv(u^2-1)(1-v^2),\\
\gamma_{DO}=(3/10){\cal{D}}{\cal{O}}(u^2-1)(5u^2-1)
(1-v^2)(1-5v^2),\\
\gamma_{QO}=-(6/5){\cal{Q}}{\cal{O}}uv(u^2-1)(1-v^2)
[(5u^2-1)(3v^2-1)+2(1-v^2)].\label{hh12}
\end{array}
\end{equation}
Here, it is explicitly seen that only the first terms
 in $\gamma_D$ and $\gamma_O$, respectively, do not
 vanish identically at $v=\pm 1$. They cancel one another
 at those points only if we use the additional condition
 (\ref{hh11a}) in the equation above, which amounts to
 put there either $\kappa =0$ or ${\cal D}+(2/5){\cal O}=0$
 plus $\kappa =1$. In the later case, $\gamma$ lastly becomes
\begin{equation}
\begin{array}{l}
2\gamma=\log(\frac{u^2-1}{u^2-v^2})+Q(u,v),\\
Q=\gamma_Q+\gamma_{DO}+\gamma_{QDO},\\
\gamma_Q=-4{\cal{Q}}u(1-v^2)-(1/2){\cal{Q}}^2(u^2-1)
(1-v^2)[u^2(9v^2-1)+1-v^2],\\
\gamma_{DO}=-2{\cal{O}}v(3u^2-1)(1-v^2)-(1/4)
{\cal{O}}^2(u^2-1)(1-v^2)\times \\
{\mbox{\hspace{1.3cm}}}(75u^4v^4 - 42u^4v^2 - 42u^2v^4 +
 18u^2v^2 + 3u^4 + 3v^4 + 1),\\
\gamma_{QDO}=-2{\cal{Q}}{\cal{O}}uv(u^2-1)(1-v^2)(3u^2-1)
(3v^2-1),\label{hh12a}
\end{array}
\end{equation}
remembering that now ${\cal D}$ is present only through
 ${\cal O}$.

It remains an arbitrariness in the full solution, namely
 the constant $\nu_0$. To show that it plays a non-trivial 
role in assuring analyticity to the solution at the black
 hole horizon, we start by writing the solution in the
 Schwarzschild coordinates $(t,R,\theta,\phi)$:
\begin{equation}
ds^2=(1-\frac{2}{R})e^{P}dt^2-e^{Q-P}[(1-\frac{2}{R})^{-1}
dR^2+R^2d\theta^2]-e^{-P}R^2 sin^2\theta d\phi^2,\label{hh13}
\end{equation}
with $P=P(u=R-1,v=cos\theta)$ and $Q=Q(u=R-1,v=
 cos\theta)$
 given respectively by (\ref{hh8}) and (\ref{hh9})
 with $\kappa=1$. At first sight we should expect that
 the singularity of this metric at the horizon remains
 only a coordinate defect since all metric functions
 here do differ from the corresponding Schwarzschild
 ones by well--behaved exponentials. To see what 
really happens there, we go to Kruskal coordinates
 $(V,U,\theta,\phi)$, defined as
\begin{equation}
\begin{array}{l}
(R-2)e^{R/2}=U^2-V^2,\\
t=2\log(\frac{U+V}{U-V}),\label{hh14}
\end{array}
\end{equation}
in order to eliminate the usual horizon divergence 
coming from the factor $(1-2/R)^{-1}$. The new,
 Kruskal components $g'_{\mu\nu}(x')$ are obtained 
from the old, Schwarzschild ones $g_{\alpha\beta}(x)$ via
\begin{equation}
g'_{\mu\nu}(x')=\frac{\partial x^{\alpha}}{\partial
 x^{'\mu}}\frac{\partial x^{\beta}}{\partial x^{'\nu}}g_{\alpha\beta}(x).\label{hh15}
\end{equation}
The line element (\ref{hh13}) reads in Kruskal's
 coordinates as
\begin{equation}
\begin{array}{l}
ds^2=F^2[-(1-H\cdot U^2)dU^2
+(1+H\cdot V^2)dV^2]\\
{\mbox{\hspace{1.0cm}}}-e^{Q-P}R^2d\theta^2-e^{-P}
R^2 sin^2\theta d\phi^2,\label{hh16}
\end{array}
\end{equation}
where $R$ (and hence $u=R-1$) is implicitly, 
analytically given in terms of $U,V$ by the first
 relation in (\ref{hh14}), $v=cos\theta$, and $F^2$
 and $H$ are defined by
\begin{equation}
\begin{array}{l}
F^2\equiv\frac{16}{R}e^{-R/2}e^P,\\
H\equiv(1-e^{Q-2P})(U^2-V^2)^{-1}.\label{hh17}
\end{array}
\end{equation}
An inspection of (\ref{hh16}) and (\ref{hh17}) clearly 
shows that this metric is analytic at the horizon
 ($u=R-1=1$ or $U=\pm V$) if and only if $H$ is analytic
 there. If we define $u=R-1=1+\epsilon$, from which
 follows that $U^2-V^2=\epsilon \; e^{(1+\epsilon/2)}$,
 this amounts just to impose the finiteness of
 $\lim_{\epsilon\rightarrow 0}H$. In fact, we see 
that $Q-2P=C_0+f(cos\theta)\epsilon+O(\epsilon^2)$, with
 $C_0$ a constant depending only upon $\nu_0$,
$\gamma_0$, ${\cal D}$, ${\cal Q}$ and ${\cal O}$, 
and $f$ being a function only of $cos\theta$, so that
 the limit
\begin{equation}
\lim_{\epsilon\rightarrow 0}H=\lim_{\epsilon\rightarrow 0}\{e^{-(1+\epsilon/2)}[\frac{(1-e^{C_0})}{\epsilon}-
e^{C_0}f(cos\theta)+O(\epsilon)]\}\label{hh18}
\end{equation}
is finite (and equals to $-e^{-1}f(cos\theta)$)
 if and only if the condition $C_0=0$ holds 
identically. This additional constraint on $\nu_0$
 and $\gamma_0$ reads
\begin{equation}
\begin{array}{l}
-2(2\nu_0+\frac{4}{3}{\cal Q})+[2\gamma_0-{\cal D}^2-
\frac{1}{2}{\cal Q}^2-\frac{21}{100}{\cal O}^2-
\frac{3}{10}{\cal D}{\cal O}]=0,\label{hh19}
\end{array}
\end{equation}
which fixes $\nu_0$. We see that the analyticity at 
the horizon and the non-existence of conical singularities
 are independent conditions, however if the necessary
condition for the absence of conical singularities
 (\ref{hh11}) also holds, (\ref{hh19}) reduces to 
the following first order relation
\begin{equation}
2\nu_0+\frac{4}{3}{\cal Q}=0. \label{hh19a}
\end{equation}
In any case both conditions taken jointly make the
 solution to be unique.

It is worth to emphasize
 that assuring analyticity to the metric at
 the horizon has nothing to do with
 removing neither strong nor conical singularities
 from curvature invariants, since the vacuum between the
 core and the shell are free from strong singularities while
the conical ones are removable from there through
 conditions (\ref{hh11}) and (\ref{hh11a}) above.
Secondly, however Kruskal coordinates 
are historically linked to
the question of geodesic completness, this issue is not
 necessary to our purposes. 
In fact, we use here the analyticity of the Kruskal
 coordinates near the horizon of the central 
black hole only to assure that
the shell is a {\it truly} controllable
 perturbation of the black hole in that region 
if expanded at any truncation order in the shell parameters,
 in the same sense considered by Vishveshwara (1970) for
 linear perturbations. Relations (\ref{hh19}) and (\ref{hh19a})
 show that analyticity near the horizon
 is a nontrivial property of the relativistic
 shells, having to be
 forced into the solution to assure its 
analytic behavior there.

\subsection{The linearized model: conical singularities
 and the Regge-Wheeler formalism}\label{RW}

We consider now the linearization of the solution with
 respect to the shell parameters, i.\ e.\ , we consider
 the shell as a small perturbation (e.\ g.\ formed by
 dust) of the central black hole geometry. Due to the
 spherical symmetry of the Schwarzschild background, 
it is assumed that {\it{any}} linear perturbation could
 be expanded in spherical harmonics, an approach due
 firstly to Regge and Wheeler (1957) to study the stability
 of black holes under small perturbations. We shall see
 that this is not true if conical singularities are 
present, in spite of the perturbation we are dealing
 with being a perfectly linear one.

We firstly summarize the Regge-Wheeler approach. The
 metric is expanded as $g_{\mu \nu}=g_{\mu \nu}^{S}+
\epsilon h_{\mu \nu}+O(\epsilon^2)$, where $g_{\mu 
\nu}^{S}$ is the Schwarzschild metric, $\epsilon$ is
 some small parameter and $h_{\mu \nu}$ is the 
general first order perturbation of the metric. We
 put $g_{\mu \nu}$ in the Einstein vacuum equations
 $\Re_{\mu \nu}=0$, where $\Re_{\mu \nu}$ is the
 Ricci tensor, and obtain $\Re_{\mu \nu}=0+\epsilon
 R_{\mu \nu}+O(\epsilon^2) =0$, which leads to the
 well known Regge-Wheeler (RW) differential equations
 $\epsilon R_{\mu \nu}=0$ for the perturbation 
$\epsilon h_{\mu \nu}$. Before trying to solve these 
equations, it is possible to expand  $\epsilon
 h_{\mu \nu}$ in tensor spherical harmonics in the
 Schwarzschild coordinates $(t,R,\theta,\phi)$
 (see Mathews (1962), Zerilli (1970) and Thorne(1980)). 
 $\epsilon h_{\mu \nu}$ falls in one of the two
 following general classes of perturbations, depending
 on its parity under rotations about the origin
 performed on the 2-dimensional manifold $t=$const.,
 $R=$const.\ : one class (the even-type one) has 
parity $(-1)^\ell$ and its general (symmetric) form is
\begin{equation}
\epsilon h_{\mu \nu}=\left(
\begin{array}{cccc}
(1-2/R)H_0^{\ell m} & H_1^{\ell m} & h_0^{\ell m}
\partial_\theta & h_0^{\ell m}\partial_\phi\\
H_1^{\ell m} & (1-2/R)^{-1}H_2^{\ell m} & h_1^{\ell m}
\partial_\theta & h_1^{\ell m}\partial_\phi\\
h_0^{\ell m}\partial_\theta & h_1^{\ell m}\partial_\theta
 & R^2(K^{\ell m}+G^{\ell m}\partial_1) & R^2
G^{\ell m}\partial_2\\
h_0^{\ell m}\partial_\phi & h_1^{\ell m}
\partial_\phi & R^2G^{\ell m}\partial_2 & R^2 sin^2
\theta (K^{\ell m}+G^{\ell m}\partial_3 )
\end{array}\right)Y_{\ell m},\label{hh28}
\end{equation}
where $Y_{\ell m}(\theta,\phi)$ is the standard 
$\ell,m$-mode spherical harmonic, $H_i^{\ell m}
(t,R)$, $h_i^{\ell m}(t,R)$, $K^{\ell m}(t,R)$
 and $G^{\ell m}(t,R)$ are the corresponding 
functions of the non-angular coordinates  and
 we define the partial differential operators $\partial_1=\partial_{\theta\theta}$,
$\partial_2=\partial_{\theta\phi}-(cos\theta /
sin\theta)\partial_\phi$, and
$\partial_3=(1/sin^2\theta)\partial_{\phi\phi}
+(cos\theta /sin\theta)\partial_\theta$.

The {\it{odd}} class has parity $(-1)^{\ell +1}$
 and its general (symmetric) form is

\begin{equation}
\epsilon h_{\mu \nu}=\left(
\begin{array}{cccc}
0 & 0 & -f_0^{\ell m}(1/sin\theta )\partial_\phi
 & f_0^{\ell m} sin\theta\partial_\theta\\
0 & 0 & -f_1^{\ell m}(1/sin\theta )\partial_\phi
 & f_1^{\ell m} sin\theta\partial_\theta\\
-f_0^{\ell m}(1/sin\theta )\partial_\phi & -
f_1^{\ell m}(1/sin\theta )\partial_\phi & 
f_2^{\ell m}\partial_4 & (1/2)f_2^{\ell m}\partial_5\\
f_0^{\ell m} sin\theta\partial_\theta &
 f_1^{\ell m} sin\theta\partial_\theta &
 (1/2)f_2^{\ell m}\partial_5 & -f_2^{\ell m}
sin^2\theta\partial_4
\end{array}\right)Y_{\ell m},\label{hh29}
\end{equation}
where $f_i^{\ell m}=f_i^{\ell m}(t,R)$ and $\partial_4=(1/sin\theta)\partial_{\theta\phi}-
(cos\theta /sin^2\theta)\partial_\phi$, and
$\partial_5=(1/sin\theta)\partial_{\phi\phi}
+cos\theta\partial_\theta-sin\theta 
\partial_{\theta\theta}$.

The superposition of perturbations are valid
 in the linearized theory, so in the matrices
 above we can assume the Einstein summation
 convention in the indices $\ell,m$ (truncated 
at the convenience) for the case of a more
 general multi-mode perturbation. We can simplify
 these matrices further by exploring the freedom
 to make arbitrary, first order in $\epsilon$, 
coordinate transformations around $(t,R,\theta,
\phi)$. In particular, the matrices achieve its 
most simple or canonical forms in the so called
 Regge-Wheeler infinitesimal gauge (Regge and 
Wheeler 1957). Nonetheless, we do not need to
 go to that gauge here.

If we linearize (\ref{hh13}) in the shell strengths
 we obtain for the perturbation $\epsilon h_{\mu \nu}$
\begin{equation}
\epsilon h_{\mu \nu}=\left(
\begin{array}{cccc}
(1-\frac{2}{R})P & 0 &  & 0\\
0 & (1-\frac{2}{R})^{-1}(-Q+P) & 0 & 0\\
0 & 0 & R^2(-Q+P) & 0\\
0 & 0 & 0 & R^2 sin^2\theta P
\end{array}\right),\label{hh30}
\end{equation}
with $P$ and $Q$ given respectively by (\ref{hh8})
 and (\ref{hh9}) after dropping the second order
 terms in the shell strengths appearing in those 
equations (remember that $u=R-1$ and $v=cos\theta$).

The linear perturbation (\ref{hh30}) is a compulsory
solution of the linearized equations $\epsilon
 R_{\mu \nu}=0$, since it is the first order term
 of the expansion of an exact solution of the full
 equations. Having in mind the axial symmetry of
 our static diagonal perturbation, an inspection 
of it shows that it is a superposition of even-type
 modes only. Then, it should be fitted by the matrix
 (\ref{hh28}) with the terms $\ell =0,1,2,3$ and 
$m=0$ retained. We find that this fitting is impossible 
in general. In fact, the linearized solution
 (\ref{hh30}) is a Regge-Wheeler perturbation
 only if it satisfy the following additional 
constraints (here $\kappa =1$):
\begin{equation}
\gamma_0=0,\;\;
{\cal{D}}+\frac{2}{5}{\cal{O}}=0.\label{hh31}
\end{equation}

On the other hand, the RW formalism lies on a 
sound mathematical basis, namely the multipole
 expansion theory (for a review with emphasis 
in general relativity, see Thorne 1980), so the
 reason for this drawback must have a physical
 origin.  In fact, as it is shown in Sokolov
 and Starobinskii (1977), to conical singularities 
there correspond
a curvature (and hence a Ricci) tensor proportional
 to a Dirac delta term centered at the symmetry
 axis. From Einstein's equations, this means
 that there correspond to them a certain
 energy-momentum tensor with this same structure
 (it does not matter here how exotic they 
should be) and hence we does not have a true 
vacuum between the core and the shell, as it
 is supposed from the starting in the RW 
formalism. By comparing (\ref{hh31}) with
 the conditions (\ref{hh11}) plus (\ref{hh11a}) 
with  $\kappa =1$ for Schwarzschild's metric,
 we see that the former amounts to rule out
 the conical singularities from the linear
 approximation. In other words, conical singularities 
{\it{survive}} to the linearization and their
 presence is an obstruction to the application of
 the RW formalism. So, what is some surprising in
 all this is that conical singularities strutting
 the model against its own gravity persist after 
the linearization, contrary to the loosely accepted
 idea that they should lie in the very non-linear
 realm of general relativity.

\subsection{The Newtonian limit of the relativistic
 models}\label{New}

We now consider the Newtonian limit of the model
 (\ref{hh13}) in the Schwarzschild coordinates 
$(t,R,\theta,\phi)$.
We assume that there exist a region
$D$ in the vacuum between the core's horizon and
 the shell
where the conditions of weak gravitational 
field and slow motion of test particles occur.
 Then, Eintein's 
equations reduce in $D$ to Laplace's
equation for the Newtonian 
potential $\Phi$, which relates to the 
metric $g_{\mu\nu}$ only through the temporal 
component
as $g_{tt}=1+(2/c^2)\Phi$. The remaining components
of the metric are irrelevant to this approximation.

Now, we consistently assume that 
$D$ is far away from the horizon, where the Schwarzschild 
coordinates
approximate to the usual time plus the
Euclidean spherical ones (we maintain the notation 
$(t,R,\theta,\phi)$ in the later approximation). Next,
 we pass to cylindrical coordinates $(t,r,z,\phi)$ via 
$z=Rcos\theta$, $r=Rsin\theta$ and expand the component 
$g_{tt}$ of (\ref{hh13}) to the first order in
$\nu_0$, ${\cal{D}}$, ${\cal{Q}}$ and ${\cal{O}}$, to 
obtain after some manipulation:
\begin{equation}
\begin{array}{l}
g_{tt}=1+(2/c^2)\Phi\equiv 1+2\nu_0-\frac{2}{R}-
\frac{4\nu_0}{R}+
2{\cal{D}}z[1-\frac{2}{R}][1-\frac{1}{R}]\\
{\mbox{\hspace{3.5cm}}}+{\cal{Q}}(2z^2-r^2)[1-
\frac{2}{R}][(1-\frac{1}{R})^2-
\frac{1}{3R^2}]\\
{\mbox{\hspace{3.5cm}}}+{\cal{O}}(2z^3-3zr^2)
[1-\frac{2}{R}][1-\frac{1}{R}][1-
\frac{2}{R}+\frac{2}{5R^2}].\label{hh21}
\end{array}
\end{equation}
The equation above was presented in VL1 (without 
the dipole) in a rather obscure (yet correct) form
for the sake of obtaining the Newtonian limit. By
 assumption, the non-dimensional constants satisfy $|\nu_0,{\cal{D}},{\cal{Q}},{\cal{O}}| \ll 1$; moreover, 
$R$ satisfies $R\gg 2$ in the region $D$, so that each
 square bracket appearing in (\ref{hh21}) reduces to
 unity in this approximation. The only place where $R$ 
itself survives is in the term $2/R$ of (\ref{hh21}), 
just that due to the monopolar core with mass $M$. The 
final step is to rewrite the surviving terms with the
 unit of length $L=GM/c^2$ appearing explicitly (remember
 that until now $R$ stood for $R/L$, etc.). The result
 for $\Phi$ is
\begin{equation}
\Phi=c^2\nu_0-\frac{GM}{R}+\frac{{\cal{D}}c^2}{L}z +\frac{{\cal{Q}}c^2}{2L^2}(2z^2-r^2)\\
+\frac{{\cal{O}}c^2}{2L^3}(2z^3-3zr^2).\label{hh22}
\end{equation}
For comparison, let us briefly recall {\it{ab initio}}
 the proper Newtonian formulation: let the coordinate
 origin stay
at the center of mass of the monopolar core (with mass
 $M$),
$z$ be the symmetry axis
of the core--shell system and $D_N$ be the region between
 the smallest and the largest spheres centered at the origin
that isolate the inner vacuum from the 
core and the shell. We have to solve
Laplace's  equation in $D_N$ for the axially
symmetric Newtonian potential.
By using the standard Legendre expansion,
we arrive at the following gravitational potential $\Phi_N$
felt by test particles evolving in $D_N$:
\begin{equation}
\Phi_N=-\frac{GM}{R}-G[I_0+I_1z
+\frac{1}{2}I_2(2z^2-r^2)
+\frac{1}{2}I_3(2z^3-3zr^2) + \cdots],\label{hh23}
\end{equation}
where $R^2=r^2+z^2=x^2+y^2+z^2$,
and $x,y,z$ are the usual Cartesian coordinates.
$I_0,I_1,I_2,I_3$ are respectively 
the constant, dipole, quadrupole and octopole shell
 strengths given by
the following volume integrals over the shell with
 mass distribution $\rho (R,\theta)$ ($P_n$ standing
 for the Legendre polynomial of order $n$):
\begin{equation}
I_n=\begin{array}[t]{c} \int\!\int\!\int \\ [-0.2cm] 
{\mbox{\tiny {shell}}} \end{array} \rho (R,\theta)\frac{P_n(cos\theta)}{R^{n+1}}dV.\label{hh24}
\end{equation}
Obviously, the regions $D$ and $D_N$ above must have
a non-empty intersection if the Newtonian approximation
 to the full relativistic case is valid, so we assume
 this and compare both expressions (\ref{hh22}) and
 (\ref{hh23}) for the potential in $D\cap D_N$, thus
 obtaining
\begin{equation}
\begin{array}{l}
|\nu_0=-\frac{m}{M}\times L\frac{I_0}{m}|\ll 1,\\
|{\cal{D}}=-\frac{m}{M}\times L^2\frac{I_1}{m}|\ll 1,\\
|{\cal{Q}}=-\frac{m}{M}\times L^3\frac{I_2}{m}|\ll 1,\\
|{\cal{O}}=-\frac{m}{M}\times L^4\frac{I_3}{m}|\ll 1,
\label{hh25}
\end{array}
\end{equation}
where $m$ is the mass of the shell.

The question is: for which Newtonian shells are these
 constraints on the constants $I_n$ attainable? To
 exemplify, let the shell be an homogeneous ring of
 mass $m$, radius $a$, and centered on the $z$-axis
 at $z=b$, whose density in cylindrical coordinates
 is $\rho=\frac{m}{2\pi a}\delta (r-a)\delta (z-b)$.
 The integrals $I_n$ in this case are
\begin{equation}
\begin{array}{l}
I_0/m=(a^2+b^2)^{-1/2}=a^{-1}(1+b^2/a^2)^{-1/2},\\
I_1/m=b(a^2+b^2)^{-3/2}=ba^{-3}(1+b^2/a^2)^{-3/2},\\
I_2/m=\frac{1}{2}(2b^2-a^2)(a^2+b^2)^{-5/2}=-\frac{1}{2}
a^{-3}(1-2b^2/a^2)(1+b^2/a^2)^{-5/2},\\
I_3/m=\frac{1}{2}(2b^3-3a^2b)(a^2+b^2)^{-7/2}=-\frac{1}{2}b
a^{-5}(3-2b^2/a^2)(1+b^2/a^2)^{-7/2}.\label{hh26}
\end{array}
\end{equation}
In the limit $a\rightarrow 0$ the ring reduces to a particle
 with mass $m$ placed at $z=b$, while if $b\rightarrow 
0$ it reduces to a ring placed at the equatorial plane
 with vanishing odd multipoles. In the case of $m/M \leq 1$,
 and given the characteristic core lenght $L=GM/c^2$
 (a relativistic parameter we know, but that can be
 crudely anticipated already in the Newtonian frame
 by assuming light can be trapped by gravity like 
ordinary matter), we see that all constraints in 
(\ref{hh25}) are satisfied for, say, $a\gg |b|,L$
 (for the limiting case of a point particle at 
$z=b$, we let $a\rightarrow 0$ and this condition 
become $|b|\gg L$). Although intuitively expected,
 this example enlightens and realizes the criteria
 to overlap both theories, that is to say, to
 assure the validity of the assumption $D\cap
 D_N\neq \emptyset$, and should be compared 
to the treatment made for example in Perry and
 Bohun (1992) for Weyl's solutions with usual,
 decreasing core-type multipoles.

We discuss now the physical role played by the
 constant $\nu_0$ in the model. Obviously,
 $\nu=\nu_0$, $\gamma=\gamma_0$ in (\ref{hh1}) 
is a solution of eqs.\ (\ref{hh2}) and (\ref{hh3})
 for any values of the constants $\nu_0$, 
$\gamma_0$. By the preceding discussion, its 
Newtonian limit leads to the constant potential
 $\Phi_N=c^2\nu_0=-GI_0$. What is the physical
 meaning of a constant Newtonian potential? It
 does describe either trivially the empty space
 or rather the interior of a very special class
 of matter distributions, the so called {\it{homoeoids}}
 (see Chandrasekhar 1987 and Binney and Tremaine 1987).
 The value of the constant potential inside an homoeoid 
is {\it{not}} arbitrary, having its value fixed by 
continuity requirements of the potential through
 the mass distribution of the homoeoid. Homoeoids 
are thus gravitationally undetectable from inside.
 This shows that the constant solution above is a 
struted, relativistic version of a Newtonian homoeoid. 
If in addition we remove the conical singularity from 
inside the relativistic homoeoid by letting $\gamma_0=0$
 (remember that conical singularities are not globally
 removable in static relativistic shell solutions since 
they are indispensable to strut the shell against its
 own gravity, so $\gamma_0=0$ really moves them outside
 the shell), we see that a simple rescaling of the time
 and the radial coordinate reduces the metric inside 
the homoeoid to that of Minkowski. On the other hand, 
the condition (\ref{hh19}) with ${\cal{D}}={\cal{Q}}=
{\cal{O}}=0$ shows that if we add a black hole inside 
an homoeoid then the metric does not extend analytically
 through the black hole horizon {\it{unless}} we maintain
 the ``strut'' in place with strength $\gamma_0=2\nu_0$. 
This is a non-intuitive aspect of the (relativistic)
 composition of a black hole with the rather simple
 homoeoidal shell.

An easy point that is worth to emphasize concerning the
 interpretation of multipolar expansions of core and shell
 types is the following: core multipoles measure
 deviations from {\it{sphericity}} of central mass
 distributions, while shell multipoles measure how 
much shells deviate from {\it{homoeoids}}, the later
 being a rather large class of distributions in which
 homogeneous spherical shells are very particular members.

We close this section with a brief discussion about
 the H\'enon-Heiles structure of the present Newtonian
 limit. It was already pointed out in VL1 that the
 potential of the shell alone pertains to the H\'enon-Heiles
 family of potentials, yet it does not suffice from its
 own to confine orbits in the intermediate vacuum (see 
the opposite signs of $r$ and $z$ in the quadrupole term 
of (\ref{hh22})). Though in practice we need to consider
 the full potential of a given galactic model viz.\ its
 effective potential $\Phi_{eff}$. The hamiltonian $H$ 
of a test particle with mass $\mu$ is given in this case by
\begin{equation}
\begin{array}{l}
H=\frac{1}{2\mu}(p_r^2+p_z^2)+\mu\Phi_{eff},\\
\Phi_{eff}=\frac{1}{2}(\frac{\ell}{\mu r})^2+\Phi_N,
\label{hh27}
\end{array}
\end{equation}
where $\ell$ is the conserved angular momentum of the
 particle associated to the axial symmetry of the
 galaxy (Binney and Tremaine 1987). It is also usual
 to assume that galaxies have further a reflection
 symmetry about an equatorial plane, in which case
 the motion restricted to that plane is integrable, 
in particular having a central stable circular orbit
 at some fixed radius, say, $r=r_0$. To study a
 nonplanar stelar orbit as a small deviation from the
 planar one, we perform a series expansion of 
$\Phi_{eff}$ in the variables $((r-r_0,z)$,  
thus approximating the full motion by a bi-dimensional
 harmonic oscillator perturbed by higher order terms. 
This is the very astronomical origin (after truncating
 the series and idealizing the numeric coefficients) of 
the cubic H\'enon-Heiles polynomial, now a paradigm of
 non-integrable potential. Some history about this 
potential in astronomy can be traced, for example,
 from Contopoulos (1960), Barbanis (1962), van de Hulst 
(1962), Ollongren (1962) and H\'enon and Heiles (1964).

By adding the other terms to the terms originated from
 the shell expansion to form $\Phi_{eff}$, we are able
 to confine test motions in the intermediate vacuum 
around the central stable orbit. If the galaxy does
 not have reflection symmetry around a middle plane 
($I_1,I_3\neq 0$ in the Newtonian case and ${\cal{D}},
{\cal{O}}\neq 0$ in the relativistic one), planar 
central stable orbits are not possible, there remaining
 only stable orbits of distorted, nonplanar type. We
 will return in section \ref{caos} to the discussion
 of the implications for orbit regularity of the
 largely assumed hypothesis of reflection symmetry
 about middle plones in galaxy modeling.

\section{Numerical results: reflection symmetry
 and chaotic motion in relativistic and Newtonian
 core-shell models}\label{caos}

We study the chaoticity of orbits in the intermediate
 vacuum of Newtonian static axially symmetric 
core-shell models and then compare them to the geodesics
 of the corresponding relativistic cases. It is 
amazing that this class of models has relativistic
 and Newtonian counterparts to be compared, there
 existing moreover sound observational motivations for both.

Specifically, we analyse in both cases the role
 played by a {\it{discrete}} symmetry, namely the
 reflection symmetry around the equatorial plane,
 in regularizing orbits against chaoticity. The
 hypothesis of reflection symmetry is a widely
 spread assumption in astronomical modeling and its
 twofold justification lies firstly in that too 
many real objects to be modeled seem in fact nearly 
symmetric with respect to a plane. On the other hand,
 the models are simplified in both analytical and
 numerical aspects by that assumption. Of course,
 we do not expect that symmetry be realized in nature
 exactly so it is relevant to search for possible
 orbital effects of its breaking. It is important
 to realize that all cases treated here preserve the 
axial symmetry --- the next continuous symmetry after
 the missing spherical one, which allows us to isolate 
the dynamical effects of only breaking or preserving
 the reflection symmetry itself.

Another advantage of the present study is that it is
 model independent (of course within the class of models
 we are dealing with) as we are focusing on general
 multipolar shell expansions instead on specific mass
 distributions.

All is made non-dimensional again in this section by
 formally taking $\Phi_N \leftrightarrow \Phi_N/c^2,
 R \leftrightarrow R/L$, etc., $H \leftrightarrow H/(\mu c^2),
 p_z \leftrightarrow p_z/(\mu c)$, etc., and $\ell 
\leftrightarrow \ell/(\mu cL)$ in equations (\ref{hh22}
--\ref{hh23}) plus (\ref{hh27}), such that the Newtonian
 system we consider comes from the hamiltonian ($R^2=r^2+z^2$)
\begin{equation}
\begin{array}{l}
H=\frac{1}{2}(p_r^2+p_z^2)+\frac{1}{2}\frac{\ell^2}{r^2}-
\frac{1}{R}\\
{\mbox{\hspace{0.5cm}}}+{\cal{D}}z+\frac{1}{2}
{\cal{Q}}(2z^2-r^2)+\frac{1}{2}{\cal{O}}(2z^3-3zr^2).
\label{hh32}
\end{array}
\end{equation}

On the other hand, the geodesic system comes from
 the lagrangean ${\cal{L}} \leftrightarrow {\cal{L}}/
(\mu c)$ given by
\begin{equation}
{\cal{L}}=\frac{1}{2}g_{\alpha\beta}\dot{x}^\alpha
\dot{x}^\beta,\label{hh33}
\end{equation}
where the metric tensor $g_{\alpha\beta}$ and the
 coordinates $x^0=t,x^1=\rho,x^2=z,x^3=\phi$ are 
obtained from the non-dimensional invariant interval
 (\ref{hh1}), and the dot stands for the derivative
 $d/ds$. The proper definition of time-like geodesics
 out of the metric interval furnishes the first 
constant of motion ${\cal{L}}=1/2$. The
 Euler-Lagrange formulation of the geodesic equations is
\begin{equation}
\frac{d}{ds}\frac{\partial {\cal{L}}}{\partial 
\dot{x}^\mu}-\frac{\partial {\cal{L}}}{\partial
 x^\mu}=0,\label{hh34}
\end{equation}
from which the two additional constants
 of motion associated to the static nature and axiality
 of the relativistic system are read out, namely
the relativistic energy $h$ and
 angular momentum $l$ defined by
\begin{equation}
\begin{array}{l}
h\equiv \frac{\partial {\cal{L}}}{\partial
 \dot{t}}=g_{tt}\dot{t},\\
l\equiv\frac{\partial {\cal{L}}}{\partial \dot{\phi}}=g_{\phi\phi}\dot{\phi}.\label{hh35}
\end{array}
\end{equation}
The remaining two equations in (\ref{hh34}) 
describe the dynamics for the variables $\rho, z$.
 We note that these $\rho, z$ are Weyl's coordinates.
 From (\ref{hh5}) we see that whenever the Schwarzschild 
coordinate $R$ satisfies $R\gg 2$ it approximates the
 usual radial spherical one (denoted here by the same
 letter $R$) and the Weyl $\rho, z$ approximate the
 usual cylindrical ones $r, z$. We remember also that 
in the Newtonian limit $dt/ds\approx 1/\sqrt{g_{tt}}$ 
(low velocities), $g_{tt}\approx 1+2\Phi_N/c^2$, so
 the relativistic energy $h$ and the Newtonian energy
 $E \leftrightarrow \mu\Phi_N/(\mu c^2)$ of an orbiting
 particle with mass $\mu$ are related through $h\approx
 1+E$, and the corresponding angular momenta through
 $l\approx \ell$. We put $\nu_0 =0$ and $\kappa =1$
 in this section. We impose the conditions (\ref{hh11}) 
and (\ref{hh11a}) for the absence of conical singularities
 on all running relativistic situations presented bellow, 
except of course for the purely dipolar shell. All
 Newtonian as well as relativistic Poincar\'e sections
 (surfaces of section) shown here are made at the
 plane $z=0$ (with the appropriate coordinate 
interpretation in each case).

In Fig.\ 1 we present typical effective Newtonian
 potential wells for the bounded orbits we are
 considering ($U$ stands for $\Phi_{eff}$ and the
 positive parts of the potential surfaces surrounding
 the wells are cut). Figs.\ 1(a) and (c) show cases 
of oblate and prolate potentials, respectively, both 
possessing reflection symmetry about the equatorial
plane given by $z=0$. Oblate cases (a) have positive
 quadrupole strengths (${\cal{Q}}>0$) while prolate
 cases (c) have ${\cal{Q}}<0$ (respectively $I_2<0$ 
and $I_2>0$, see eqs.\ (\ref{hh22}) and (\ref{hh23})
 and, for the specific case of a ring, eq.\ (\ref{hh26})).
 The presence of the reflection symmetry implies the 
vanishing of all odd shell multipoles (${\cal{D}}$, 
${\cal{O}}$ and $I_1$, $I_3$ here). We see that the
 oblate case has only one unstable equilibrium point
 (in the equatorial plane) while the prolate case has 
two of them (symmetrically placed outside the equatorial
 plane), which is easy to understand in terms of the
 exterior oblate/prolate shell mass distributions. Figs.\ 
1 (b) and (d) show the typical deformation of the previous
 oblate and prolate cases, respectively, when we break
 the symmetry of reflection about $z=0$ by introducing
 non-vanishing odd shell multipoles (${\cal{D}}$, ${\cal{O}}$).

On the other hand, the relativistic regime is felt in 
the situation we are concerned with mainly through the
 existence of one more unstable equatorial equilibrium 
point in addition to those already present in the 
Newtonian potential. 
This intrinsically relativistic additional unstable point
 is associated to the presence of the black hole at the
 center and marks the point above which the orbits fall
 into the black hole. In the remaining, the relativistic
 potentials are qualitatively similar to those displayed
 in Fig.\ 1 for the Newtonian case.

We have made a wider exploration of the shell parameters,
energy and angular momentum
than shown here, with the same conclusions in
 large. The parameter values we actually chose to show
 are a compromise between
 the need to make consistent comparisons
 and the sharpening of the effects we found.

The case of a purely dipolar shell has been anticipated
 in VL2, where we stressed that the monopolar core plus
 shell dipole is non-trivial already in the Newtonian
 context. Moreover, we show that the Newtonian case is
 integrable whereas the relativistic one is chaotic,
 which justifies the characterization of chaos in the
 purely dipolar case as an intrinsic general relativistic
 effect. Here, we explore the fact that the Newtonian 
dipolar shell breaks the reflection symmetry of the model
 without breaking the integrability of the motion itself. 
Thus core-shell models provide us with two very distinct
 integrable situations from the point of view of the 
reflection symmetry: core plus dipolar shells (D-cases),
 which do not have that symmetry, and purely monopolar
 cores (Kleperian cases) which do have it. A typical
 Poincar\'e section of an integrable D-case is shown 
in Fig.\ 2(a).

Firstly, we perturb both integrable configurations
 above with a reflection symmetry preserving,
 {\it{oblate}} quadrupolar term (${\cal{Q}}>0$).
 Surfaces of section for core plus oblate quadrupolar
 shells (oblate Q-cases) are shown in Figs.\ 2(b,c,f)
 while sections for core plus oblate dipolar-quadrupolar
 shells (oblate DQ-cases) are shown 
in Figs.\ 2(d,e). These figures show that the breaking of
 the integrability of the D-cases (without reflection 
symmetry) by a quadrupole is much stronger than that
 of the Keplerian case (with reflection symmetry) by
 the same quadrupole. Moreover: Fig.\ 2(c) shows that
 chaoticity in the oblate Q-cases is in fact present
 only in a residual, ``microscopic'' level, Fig.\ 2(f)
 illustrates the robustness of the strong regularity of
 the oblate Q-cases against the varying of the quadrupole 
strength, while Fig.\ 2(e) shows that the strong chaoticity
 in the DQ-cases is very dependent on the energy, in 
contrast with the robustness of the almost regularity
 of the oblate Q-cases also with the energy.

Fig.\ 3 shows the effects of introducing a reflection
 symmetry breaking, octopolar perturbation on the two 
integrable configurations (D-cases and Keplerian cases),
 and also on the almost regular oblate Q-cases. Figs.\ 3(a)
 and (f) show the effects of a purely octopolar shell on
 the Keplerian case (O-case) at two different octopole 
strengths. We see that these effects are very much stronger
 in comparison with the almost-regular oblate Q-cases. Figs.\
 3(b-e) show the effects of the various combinations of shell
 multipoles and confirm the dominance of the chaotic effects
 associated to the odd multipoles over those related to the
 purely oblate Q-cases. In particular, the enhancement of
 chaoticity associated to the breakdown of the reflection
 symmetry is reinforced by the oblate QO-cases presented 
in Fig.\ 3(c) in comparison with the oblate Q-cases, and
 Figs.\ 3(d) and (e) confirm the strong dependence with 
the energy of the chaoticity associated to the lack of
 reflection symmetry.

Fig.\ 4 shows surfaces of section for {\it{prolate}} 
quadrupolar cases (${\cal{Q}}<0$). Figs.\ 4(a-c) show
 the very much higher chaoticity of the prolate Q-cases
 in comparison to the previous quasi-regular oblate
 Q-cases. This is to be expected in view of the prolate 
Q-cases having two unstable equilibrium points (and hence
 suffering the simultaneous influence of the two
 corresponding instability regions) instead of only
 one occurring in the oblate Q-cases. These figures
 also show that the central primary stable orbit,
 deeply inside the accessible region, is strikingly
 the first one to bifurcate and drive the chaotic
 behavior with the rising of the energy. The strong
 dependence with the energy of the whole set of
 bifurcations is also a distinctive aspect of these
 figures. In Fig.\ 4(d) we show the effect of applying
 the perturbing prolate quadrupolar term on the
 integrable, reflection symmetry missing D-cases
 (prolate DQ-cases). We find that the absence of
 mirror symmetry in the prolate DQ-cases causes 
the opposite to that occurring in the oblate DQ-cases,
 namely the missing mirror symmetry does greatly enhance
 the orbit regularity in the prolate cases. Fig.\ 4(e) 
does confirm that the breakdown of mirror symmetry
 associated to the addition of an octopolar term to 
the prolate Q-cases (prolate QO-cases) strongly reduces
 the chaotic manifestation of the prolate Q-cases
 themselves. In particular, Figs.\ 4(d-f) show evidences
 that the lack of symmetry introduces a robust family
 of regular asymmetric orbits around a stable primary
 one in the bounded region. This regularizing effect 
due to the lack of mirror symmetry in the prolate DQO-cases
 is easily understandable in terms of the progressive
 lack of influence of one among the two unstable
 equilibrium points of the potential on the bounded 
region accessible to the particle if odd shell multipoles
 are put in scene (cf.\ Figs.\ 1(c) and (d)).

We present in Fig.\ 5 the findings for the
 corresponding {\it{relativistic}} core-shell
 models. We choose for these figures the same
 non-dimensional multipole shell strengths and
 angular momenta of the corresponding Newtonian
 figures and, except for the purely dipolar shell,
 we always eliminate the conical singularities (or 
``struts'') from the intermediate vacuum by fixing 
$\gamma_0$ and ${\cal D}$ in terms of ${\cal Q}$
 and ${\cal O}$ through the conditions (\ref{hh11})
 and (\ref{hh11a}). Also, we always put $\nu_0=0$
 here. The relativistic energy $h$ is chosen in each 
case to be close that given by the approximate relation
 $h\approx 1+E$ where $E$ is the Newtonian energy.

Fig.\ 5(a) shows a typical surface of section for a black
 hole plus purely dipolar shells (RD-cases), its strong
 chaoticity having to be compared with both the almost
 regularity of the {\it{oblate}} relativistic quadrupolar 
case (RQ-case) shown in Fig.\ 5(b) (its very small chaotic
 zones are visible if we zoom the cross-type regions
 between the islands) and the corresponding integrable 
Newtonian D-case shown in Fig.\ 2(a). [In particular,
 the smallness of the chaotic zones of the oblate RQ-case
 shown in VL1 occasioned the misleading statement made by
 us therein of these cases not exhibiting chaos, which
 was promptly corrected in Vieira and Letelier 1996b.]

In fact, Fig.\ 5(b) does not represent the full chaotic
 behavior possible to oblate RQ-cases. As we saw above
 about the relativistic counterparts of the Newtonian 
potentials, the relativistic oblate cases have one more
 unstable scape point (in fact, an infall point into 
the black hole) inermost at the equator in addition
 to that one provided in both cases by the oblate shell
 itself. Hence, if we vary the energy and the angular
 momentum so that both unstable regions are equally 
accessible by the bound nearly equatorial geodesics, 
it is possible to enhance strongly the chaotic behavior
 of Fig.\ 5(b) to encompass even a significant portion
 of the more external region of the surface of section
 (that associated to the nearly equatorial orbits). We
 will see more about this enhancement for the relativistic
 cases below in Fig.\ 5(e).

Fig.\ 5(c) shows a surface of section for relativistic
 dipolar-octopolar cases (RDO-cases). Remember that 
both odd components are simultaneously needed in view
 of the off-strut constraint (\ref{hh11a}). Fig.\ 5(d)
 is the same for full oblate RDQO-cases. Figs.\ 5(a-d)
 show that the chaotic zones are greatly enlarged whenever
 odd shell multipoles are present, in contrast with the
 almost-regular oblate RQ-case shown.

Figs. 5(e) shows a surface of section for relativistic
 {\it{prolate}} cases (prolate RQ-cases). It is to be
 compared to the corresponding prolate Newtonian one
(see Fig.\ 4(b)), in particular we note the evident
 fingerprints of the bifurcation series starting from 
the originally stable central primary orbit already
 present in the prolate Q-cases. We also see that the 
prolate RQ-cases share with the prolate Q-cases the
 same relatively wild chaotic behavior as compared 
to the respective oblate (R)Q-cases.

Like in the Fig.\ 5(b), the dynamics would be more
 chaotic in Fig.\ 5(e) if the angular momentum and
 the energy (unaltered here for the sake of full
 comparison) were ajusted to allow all the three
 unstable points of the relativistic potential 
(instead only the two ones related to the prolate
 character of the shell, see Fig.\ 1 and the comments 
therein about the additional black hole unstable point) 
have stronger combined influence on the motion. So, the
 remaining differences  between the relativistic figure
 5(e) and the Newtonian one 4(b) are due to the following
: the central region (in the surface of section) is the only
 one able to become chaotic in the Newtonian case since
 its orbits are just those (non-planar ones) that can
 reach the unstable regions near the two unstable scape
 points symmetrically placed outside the equatorial plane,
 while the more external (in the surface of section)
 quasi-equatorial orbits are regular since they are far
 from those two points and under the strong influence of
 the equatorial stable orbit in the boundary of the surface
 of section. When we change the Newtonian central mass by
 a true black hole then a third unstable scape point (in
 fact a rolling down point into the black hole) is placed
 innermost in the equatorial plane. Associated to it there
 is one more unstable region which can be reached, this
 time by the quasi-equatorial orbits. The specific combination
 of parameters of Fig.\ 5(e) is such that the central,
 non-planar orbits can feel only moderately the influence
 of the two scape points linked to the presence of the 
prolate shell, while the quasi-equatorial orbits are feeling
 strongly the nearby influence of the unstable point 
associated to the black hole. As we said above, other
 combinations of parameters will allow a much stronger
 spread and eventually the overlaping of both types of
 chaotic regions.

Finally, Fig.\ 5(f) shows the restoration of a robust
 set of regular asymmetric orbits around a primary 
stable one when we break the reflection symmetry 
around the equatorial plane by switching on the odd
 multipoles onto the prolate RQ-cases, much the same 
as occured in the prolate Newtonian Q-cases.

\section{Discussion}\label{discussion}

In the first part of this article we presented an
 unifying discussion concerning with the properties
 and, more important, the physical content of some
 relativistic, static, axially symmetric core-shell
 models on their own as well as in connection with
 the corresponding linearized and Newtonian models
 taken as limiting cases. The analytical and observational
 motivations for these relativistic and Newtonian models
 were also shown. This was acomplished in a reasonably 
self-contained manner in the introduction and section 
2 and needs no additional comments.

In the second, numerical part of this work we explored
 the fact that the models i) are generic within the
 class they pertain and ii) have exact relativistic 
and Newtonian counterparts to be compared, to study
 the chaoticity of bound orbits in the vacuum between 
the core and the shell.

Specificaly, we firstly tested the relevance of the
 presence/absence of the reflection symmetry around
 the equatorial plane for the chaotic behavior of 
the orbits. We find consistent evidences for a
 non-trivial role played by the reflection symmetry 
on the chaoticity of the dynamics in both the
 relativistic and Newtonian cases. We summarize these 
findings as follows: the {\it{breakdown}} of the
 reflection symmetry about the equatorial plane in
 both Newtonian and relativistic core-shell models 
does i) {\it{enhance}} in a significant way the 
chaoticity of orbits in
 reflection symmetric {\it{oblate}} shell models
 and ii) {\it{inhibit}} significantly also the 
occurrence of chaos in
 reflection symmetric {\it{prolate}} shell models.
 In particular, the lack of the reflection symmetry
 provides the phase space in the prolate case with
 a robust family of regular orbits around a stable
 periodic orbit that is otherwise missing at higher
 energies.

The other point we addressed about the chaotic behavior
 of orbits was for the consequences of substituting
 true central black holes by Newtonian central
 masses. We find that the relative extents of the
 chaotic regions in the relativistic cases are 
significantly larger than in the corresponding
 Newtonian ones. Although not surprising in 
thesis, the strong differences between both 
regimes are in order in view of the procedure 
found in the literature of simulating the presence
 of a black hole at the core of a galaxy with a
 naive $-1/r$ Newtonian term (see for example 
Gerhard and Binney 1985, Sridhar and Touma 1997
 and Valluri and Merrit 1998). This approximation 
is certainly valid far from the black hole but not
 in its proximity as is the case here.

These findings stress i) the non-trivial role
 of the reflection simmetry in both relativistic
 and Newtonian regimes, in contrast with its 
universal acceptance in astronomical modeling,
 ii) the strong qualitative and quantitative 
differences between relativistic and Newtonian
 regimes, in particular when dealing with orbits 
in the vicinity of black holes, and iii) the 
intrincate interplay between both aspects when 
they are simultaneously present.

The true dynamical aspect related to the role of
 reflection symmetries on the regularity of the
 orbits refers of course to the parity of the
 constants of motion (the famous third one in 
the present axially symmetric case) with respect
 to the coordinates, as well as their power to 
regularize orbits in phase space. Translated to
 these terms, the findings above are saying that 
the additional constant of motion in question is
 by far more powerful whenever i) it is an 
{\it{even}} function of the coordinates for
 {\it{oblate}} cases, or ii) it is an {\it{odd}}
 function of the coordinates for {\it{prolate}}
 cases. Since the issue of finding global or even
 approximate constants of motion given the dynamics
 is hard if not feasible in most cases, the former
 procedure of simply checking the presence/absence 
of reflection symmetries is of much more practical interest.

Additional study is needed to see if and how our
 findings are extendible to more realistic configurations,
 with and without black holes/central masses.
 The obvious improvement is to fill the intermediate
 vacuum with some reasonable mass distribution.
 While this is far from obvious in general relativity,
 it is easy to do in the Newtonian context. For
 example, we are considering just superpose our
 shell multipoles to some relevant potentials of
 celestial mechanics such as Plummer-Kuzmin, Ferrers
 and others and even triaxial potentials. They have
 one or more planes of symmetry, with or without 
axial symmetry, and so we can perturb them with
 shell terms to verify similar effects to those we found.

Here we considered only tube orbits ($l_z\neq 0$)
 since we are interested in the effects of the 
existence of planes of symmetry on {\it{tridimensional}}
 motion (somewhat similar effects to the found here are
 seen in Gerhard 1985 for the restricted case of planar
 motions with respect to the existence of one or more
 lines of symmetry). The model improvements above will
 allow us to study the effects of breaking the reflection 
symmetry also on tridimensional box orbits.

Meantime, the fully relativistic program is
 in progress. We have recently succeeded in
 giving rotation to a black hole (i.\ e.\ in
 converting it into a Kerr black hole) plus
 a dipolar shell term (Letelier and Vieira 
1997). There are increasing evidences for the
 existence of black holes, particularly inside
active galactic nuclei (Kormendy and Richstone
 1995), which motivate us to consider also
 rotating core-shell models in the same lines
 followed here.

\acknowledgments

The authors thank FAPESP and CNPq for financial
 support, and Jorge E. Horvath and Andre L. B.
 Ribeiro for discussions.

\clearpage

\figcaption{Typical effective Newtonian potential
 wells we are dealing with. $U$ stands for $\Phi_{eff}$
 and the positive parts of the full potential
 surfaces are cut. (a) shows an even (${\cal{D}}
={\cal{O}}=0$) oblate (${\cal{Q}}>0$) shell and
 (b) shows its perturbation due to the presence
 of the odd shell multipoles ${\cal{D}}$ and/or
 ${\cal{O}}$. On the other hand, (c) shows an
 even prolate (${\cal{Q}}<0$) shell and (d) shows
 its perturbation due to the presence of the odd 
shell multipoles. Note that (a) and (c) have a
 plane of symmetry, which is broken by the odd 
multipoles in (b) and (d).}

\figcaption{This figure exhibits, for the parameter
 values shown, surfaces of section at the plane
 $z=0$ of integrable Newtonian configurations
 (D-cases and Keplerian cases) perturbed by
 {\it{oblate}} shell quadrupoles only. Note
 that quadrupolar perturbations do preserve the 
mirror symmetry. In all figures L accounts for 
the non-dimensional angular momentum $\ell$. (a)
 shows a typical section of the integrable D-case.
 This case does not have a plane of symmetry, in 
contrast with the also integrable spherically 
symmetric Keplerian case. (b), (c) and (f) show
 the perturbed Keplerian cases (oblate Q-cases), 
while (d,e) show perturbed D-cases (DQ-cases).
 Note that orbit regularity is strongly broken
 (preserved) in the absence (presence) of mirror
 symmetry, this conclusion being robust against
 multipole strength variations.}

\figcaption{In this figure we add octopolar
 components (${\cal{O}}\neq 0$) to the shell,
 which necessarily does break the mirror
 symmetry. In (a) and (f) we see the strong
 chaotic effect of perturbing Keplerian cases
 with octopoles (O-cases) (compare with the
 almost-regularity of Figs.\ 1(b,c,f)). In (b)
 we see the  strong chaotic effect of octopoles
 on integrable D-cases (DO-cases), which is 
fully expected as mirror symmetry is already
 absent from the starting. In (c) we break the
 reflection symmetry of the almost-regular 
oblate Q-case with octopoles (QO-cases) and
 in (d,e) we present the full DQO-case for two
 slightly different energies, which confirms
 the strong dependence of the chaoticity with 
the energy when  the mirror symmetry is broken, 
in contrast with the robustness of the almost-regular,
 mirror symmetric cases against energy variations.}

\figcaption{We show in this figure {\it{prolate}}
 Q-cases. In (a-c) we illustrates the well known 
fact that (mirror symmetric) prolate Q-cases are
 strongly chaotic on their own and that their
 chaoticity is highly energy dependent. We note
 in particular that the orbit bifurcations toward
 the chaotic behavior unusually start from the
 central primary stable orbit itself rather than
 from the boundary. In (d-f) we break the mirror
 symmetry of the prolate Q-cases with different
 combinations of odd multipoles. This does cause
 a strong regularizing effect on the orbits, in
 particular by restoring, of course in a distorted 
fashion, the whole family of regular orbits around
 a primary stable orbit. Moreover, this restoration
 is robust against multipole strength as well as
 energy variations.}

\figcaption{We present some surfaces of section
 at the plane $z=0$ for the corresponding fully
 relativistic core-shell configurations. In this
 figure, the coordinates $r,z$ stand for the
 Weyl ones $\rho$, $z$ (see text for more details).
 In addition, in (a) we set $\nu_0=\gamma_0=0$ (in
 the purely dipolar case conical singularities 
(CSs) are unavoidable), while in (b-f) we set
 $\nu_0=0$ and, in order to eliminate CSs,
 $\gamma_0$ and ${\cal D}$ are given in terms
 of ${\cal Q}$ and ${\cal O}$ according to the
 conditions (\protect\ref{hh11}) and
 (\protect\ref{hh11a}). The relativistic
 case confirm the role of the presence/absence
 of the reflection symmetry on the chaoticity 
of the orbits already detected in the preceding 
Newtonian figures.}

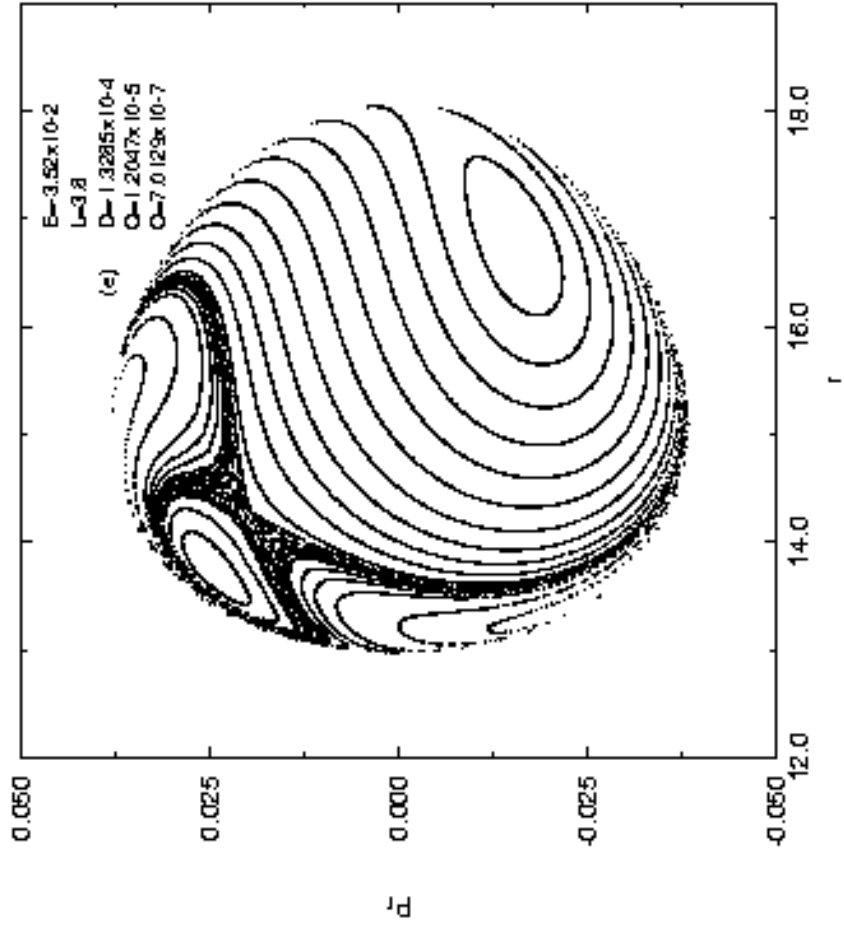
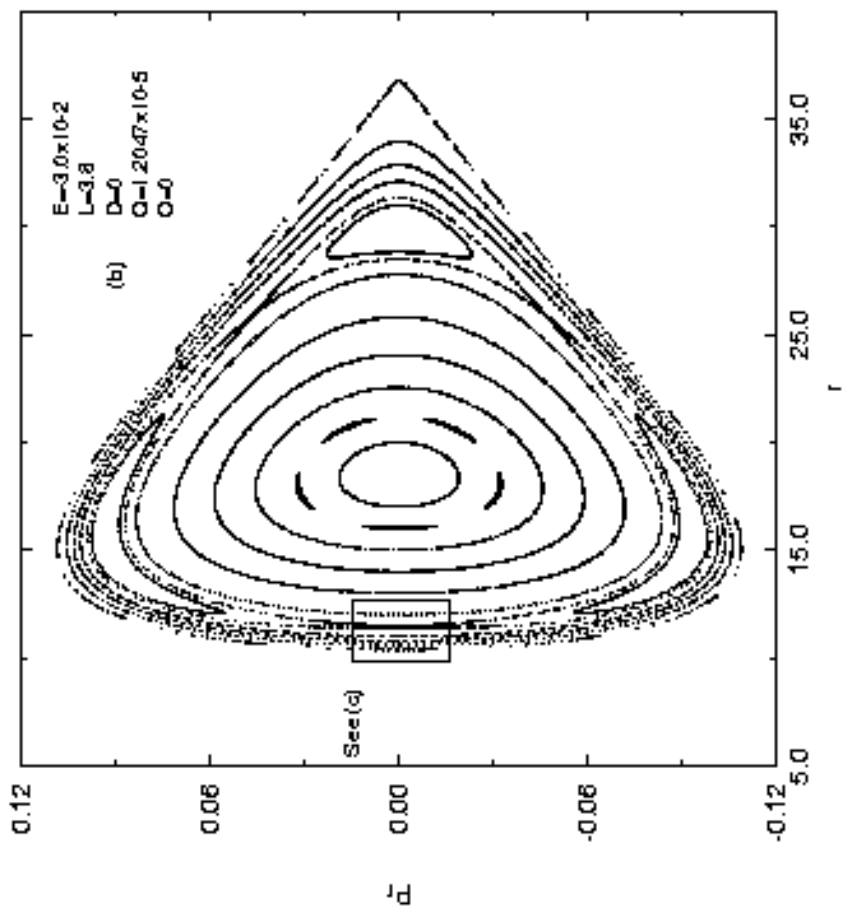
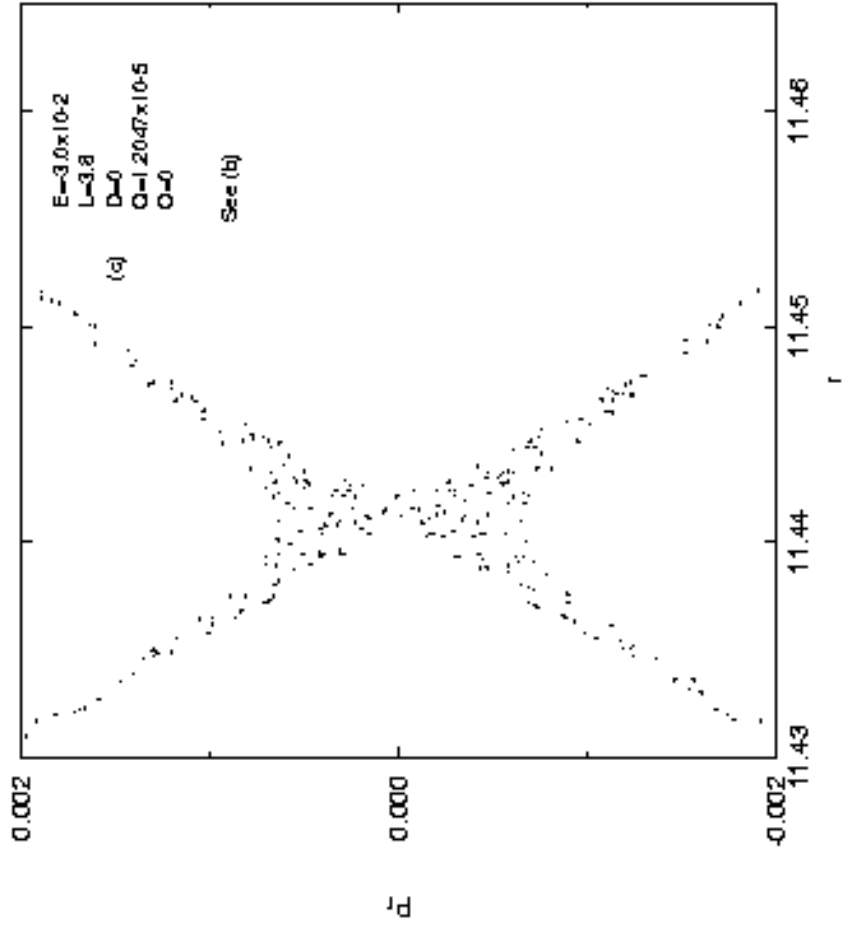
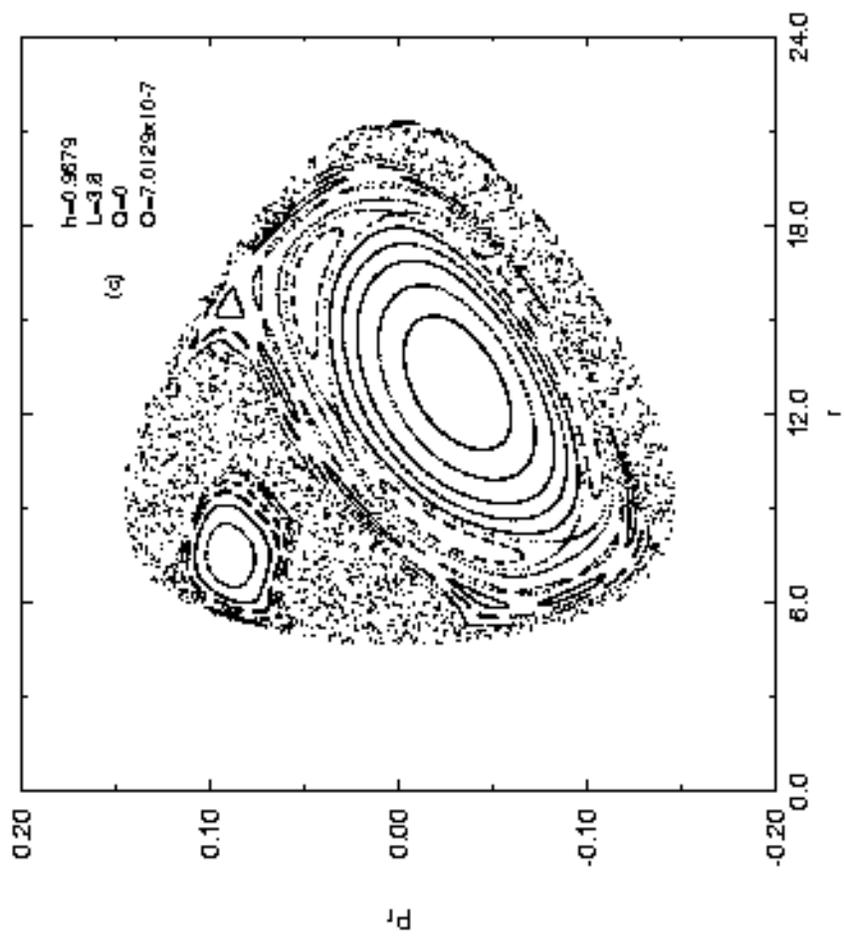
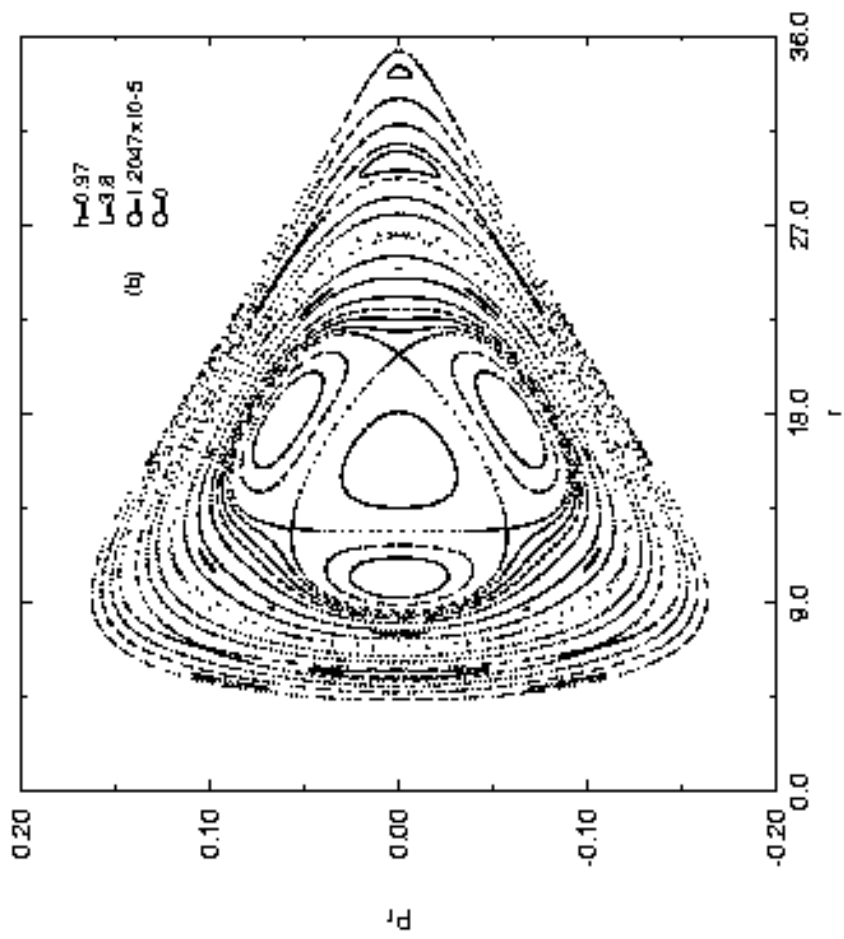
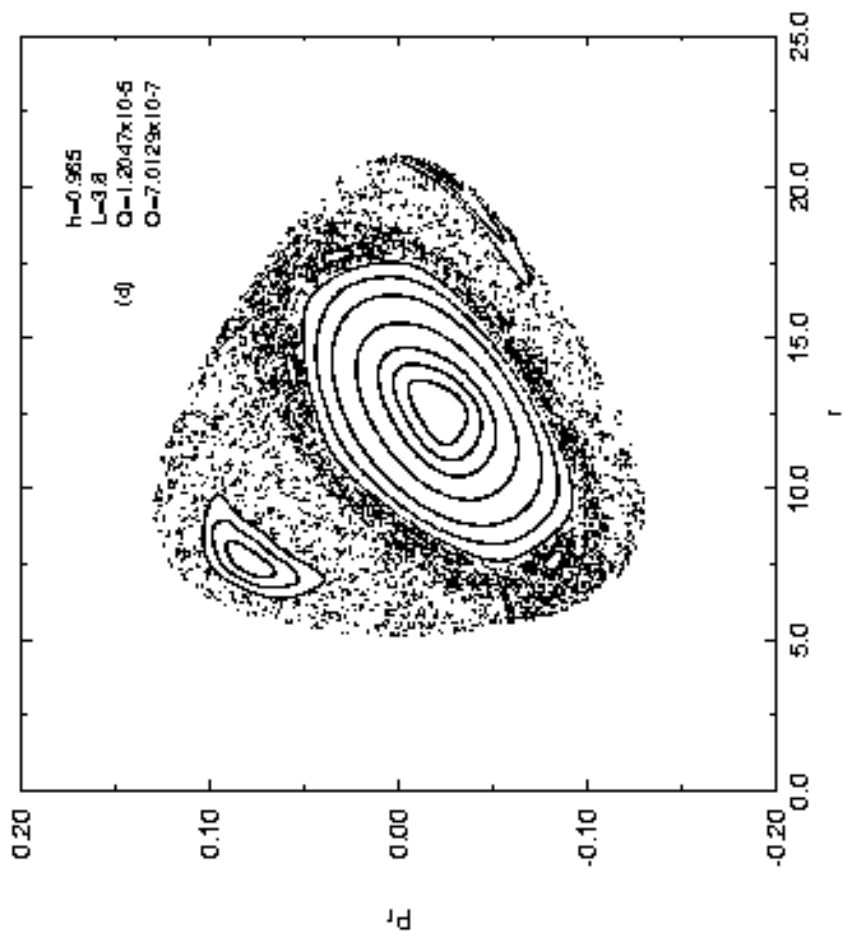
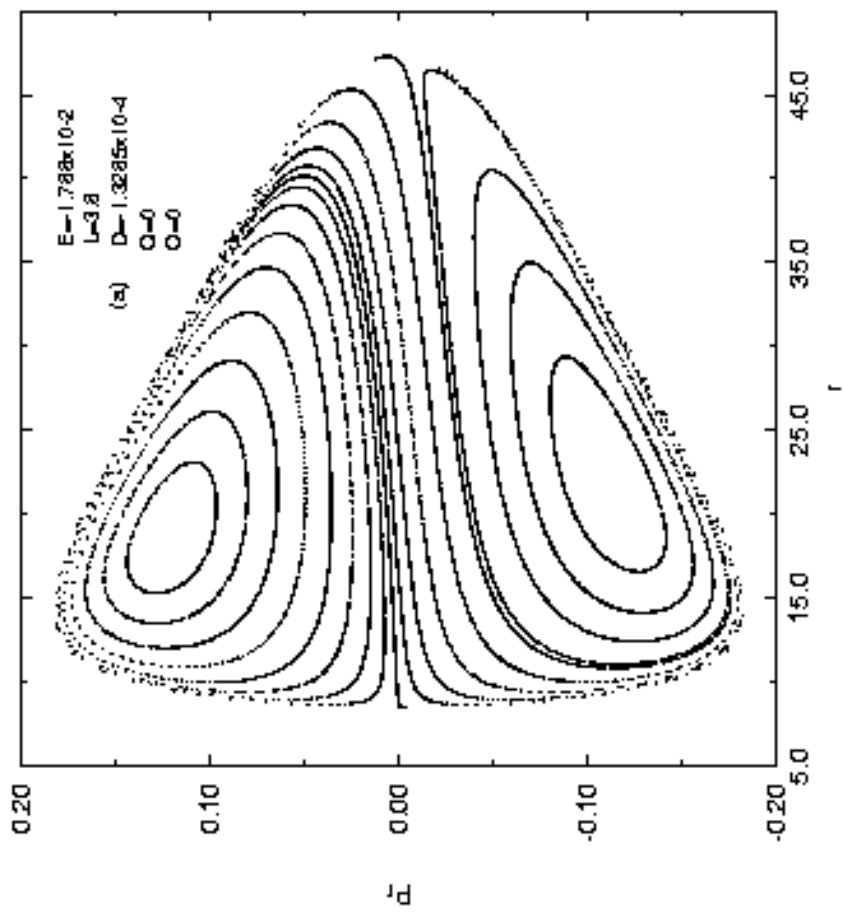
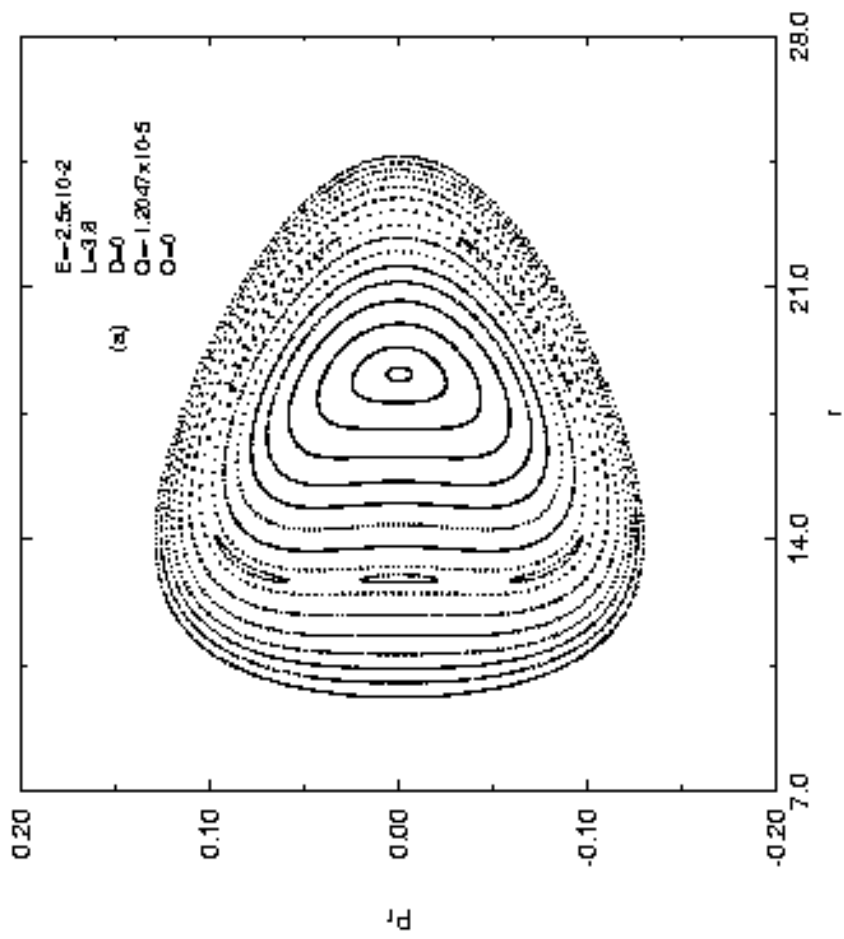
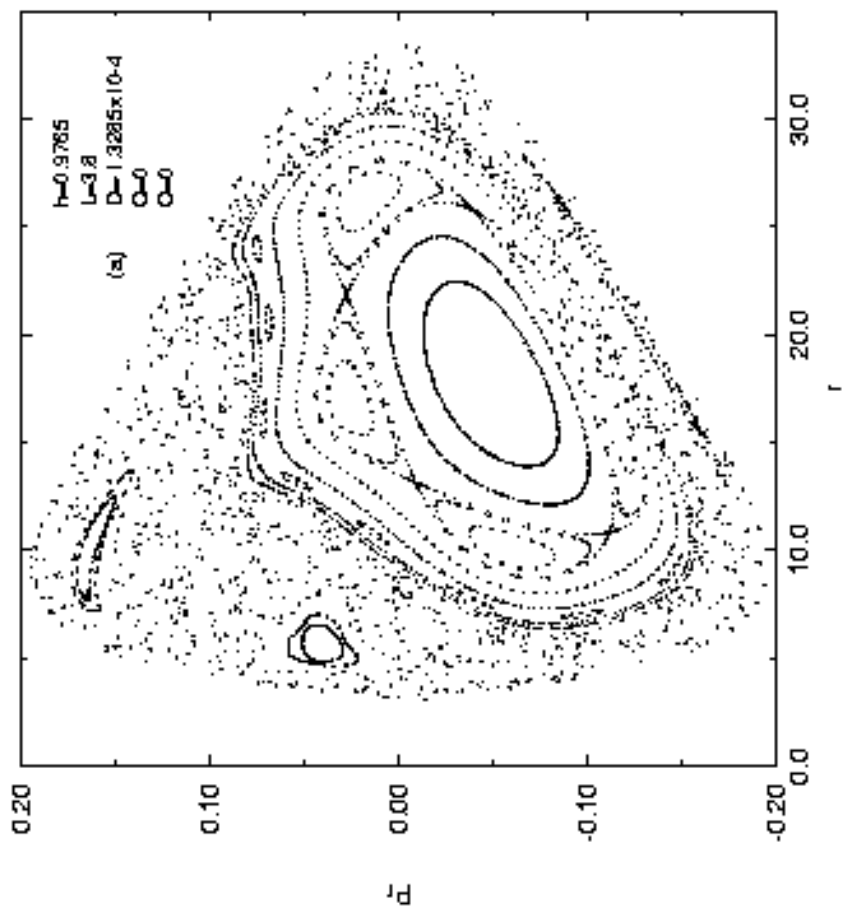
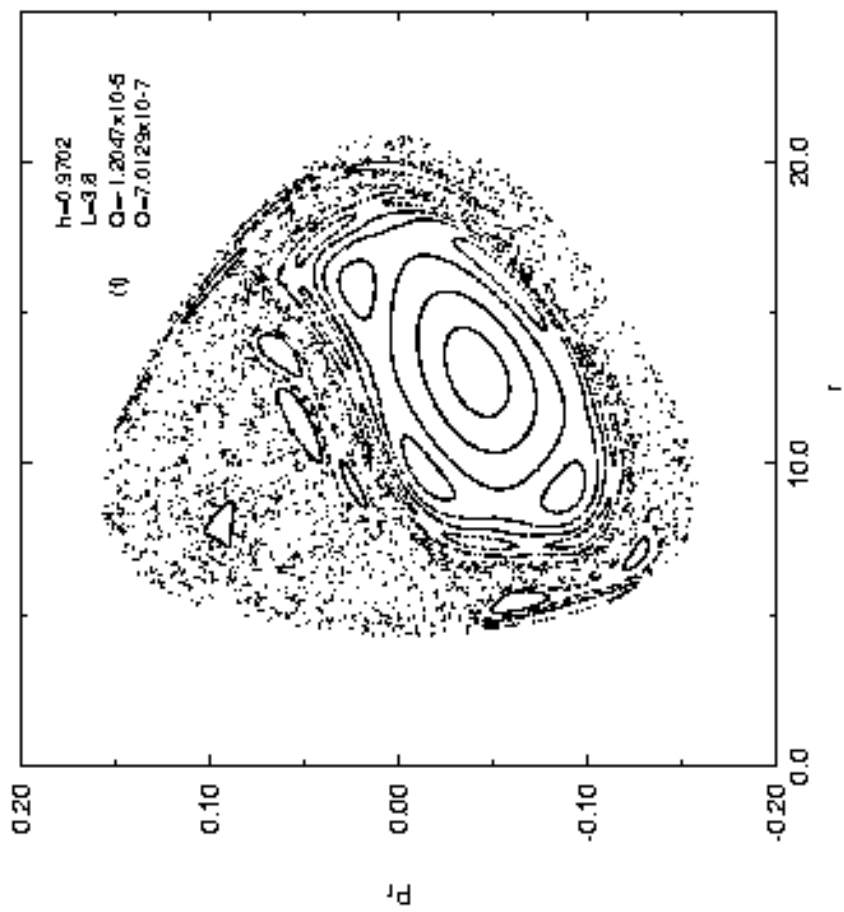
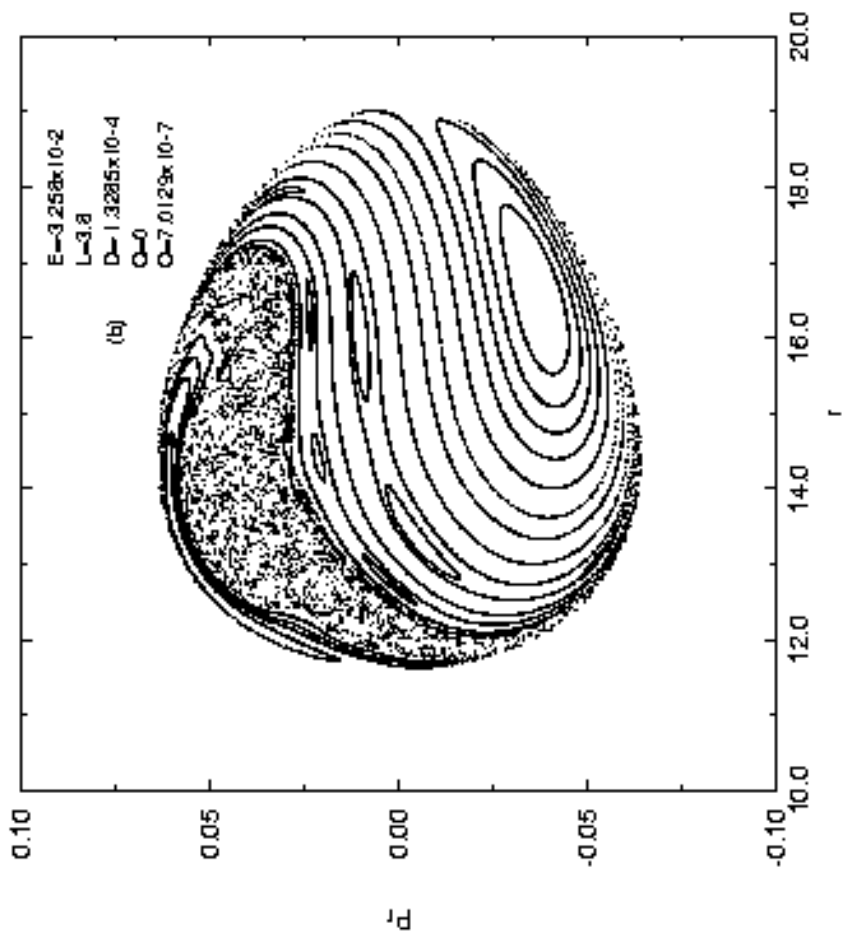
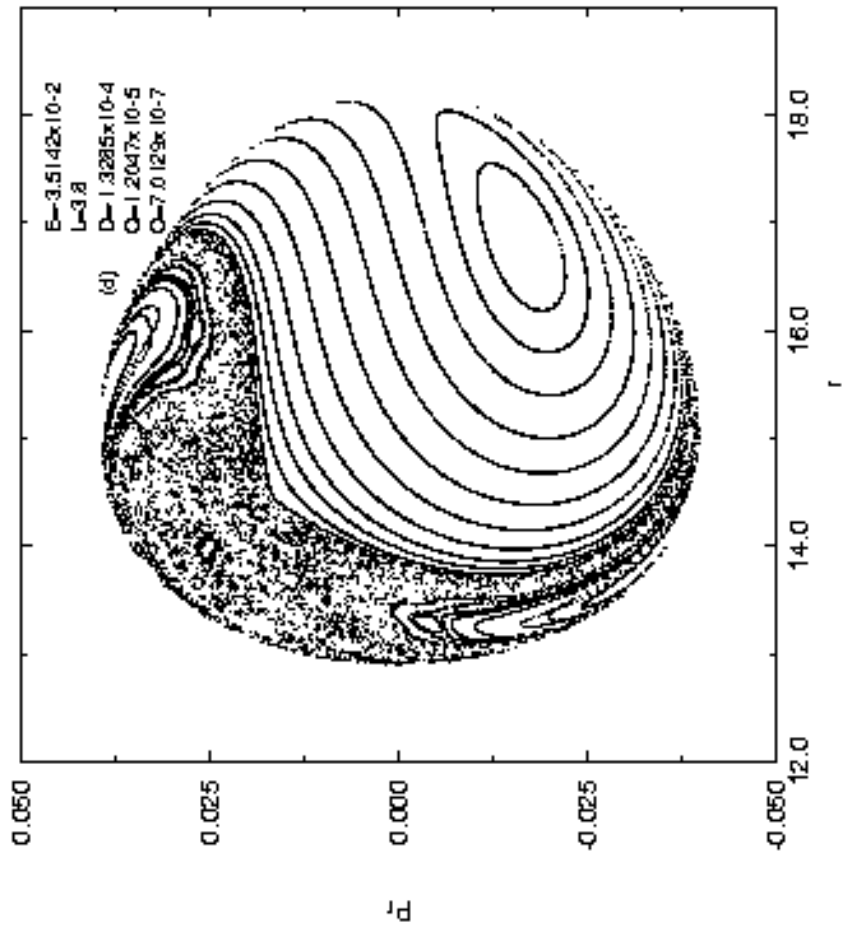
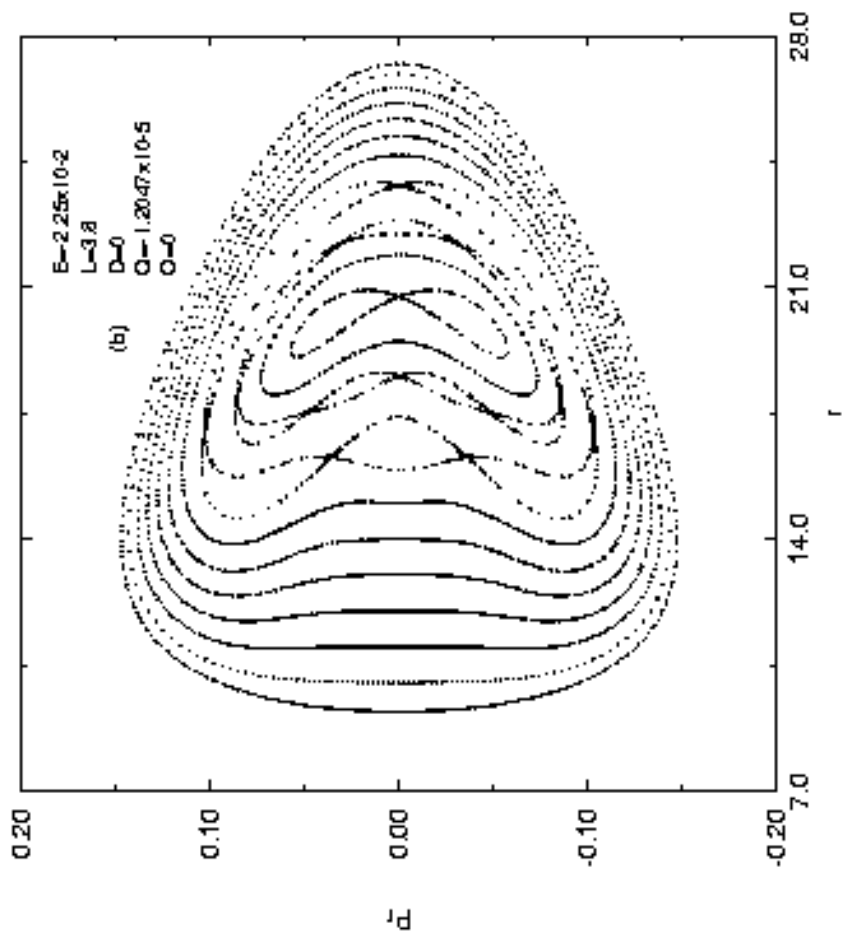
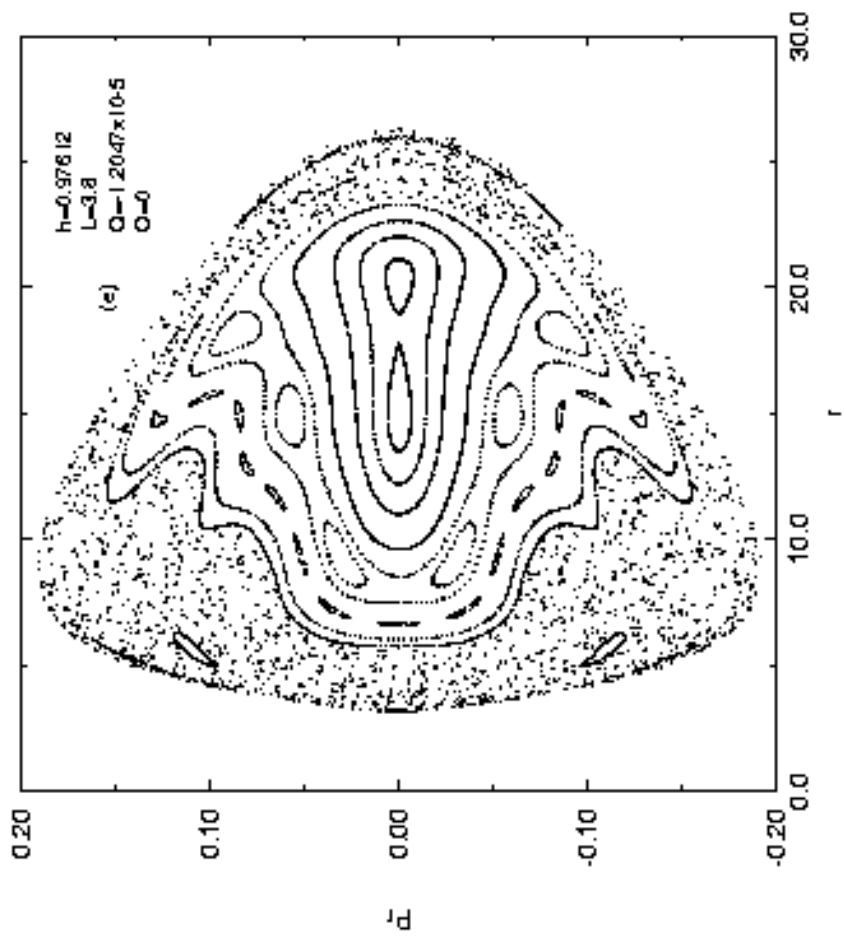
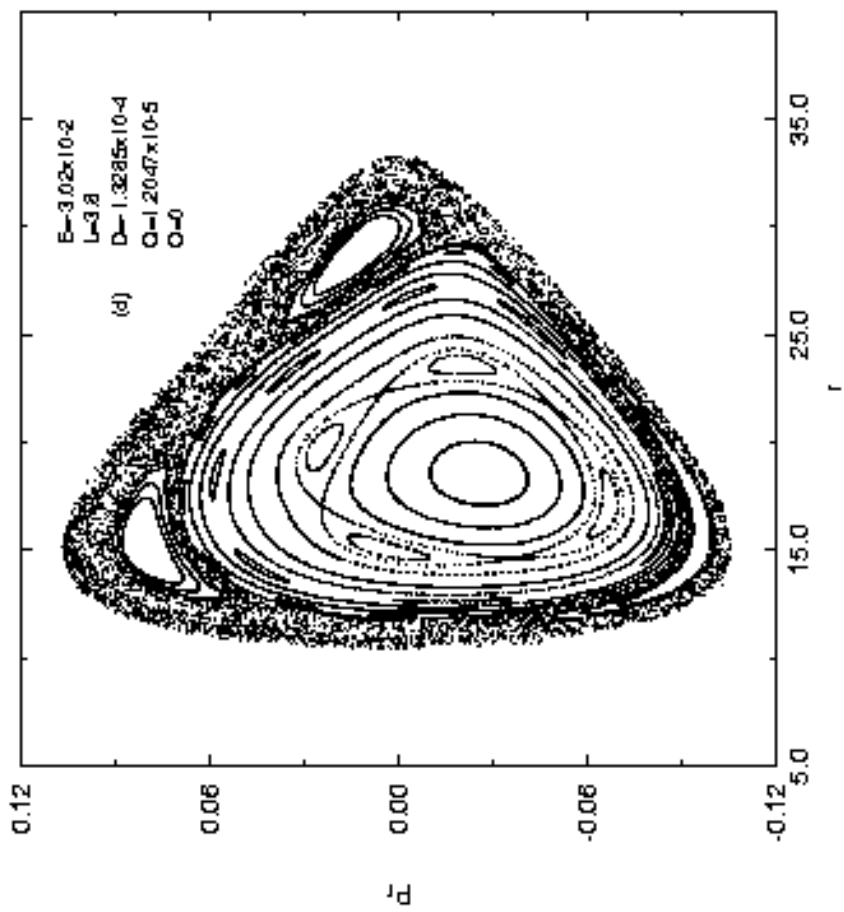
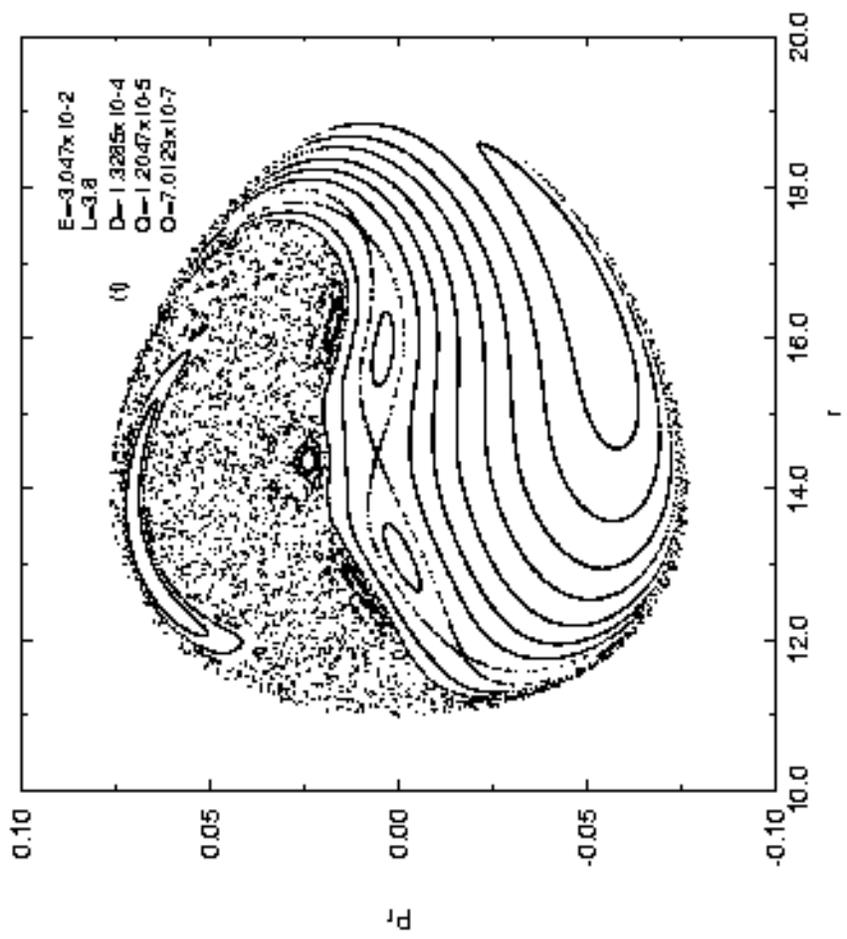
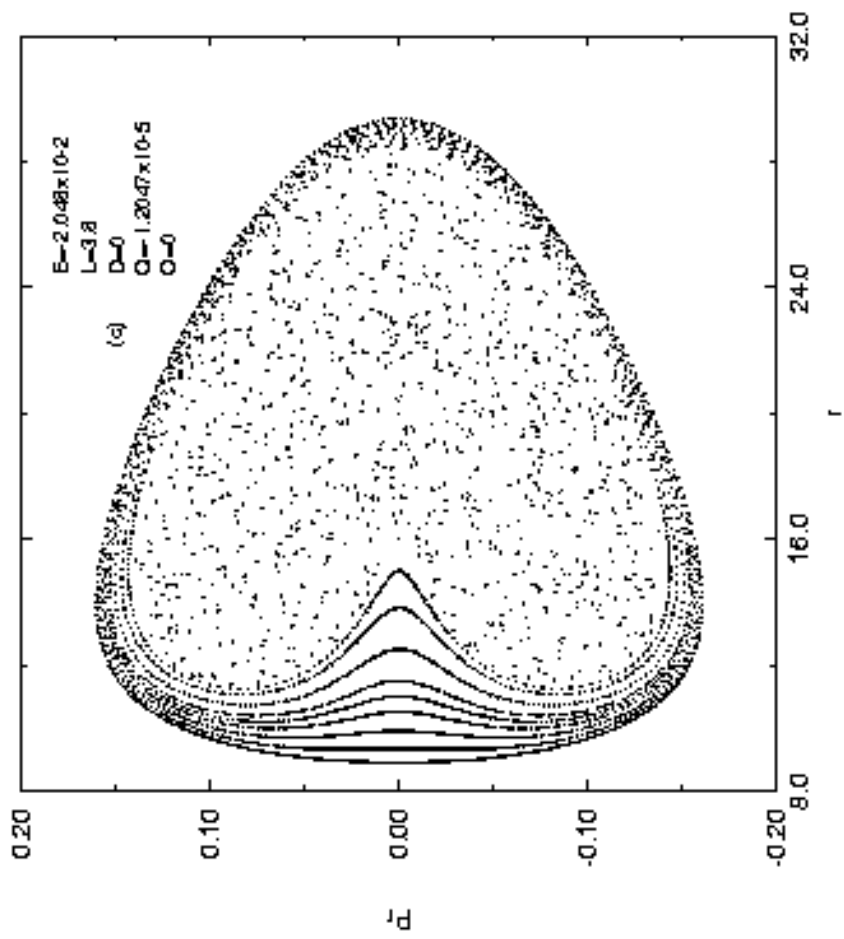
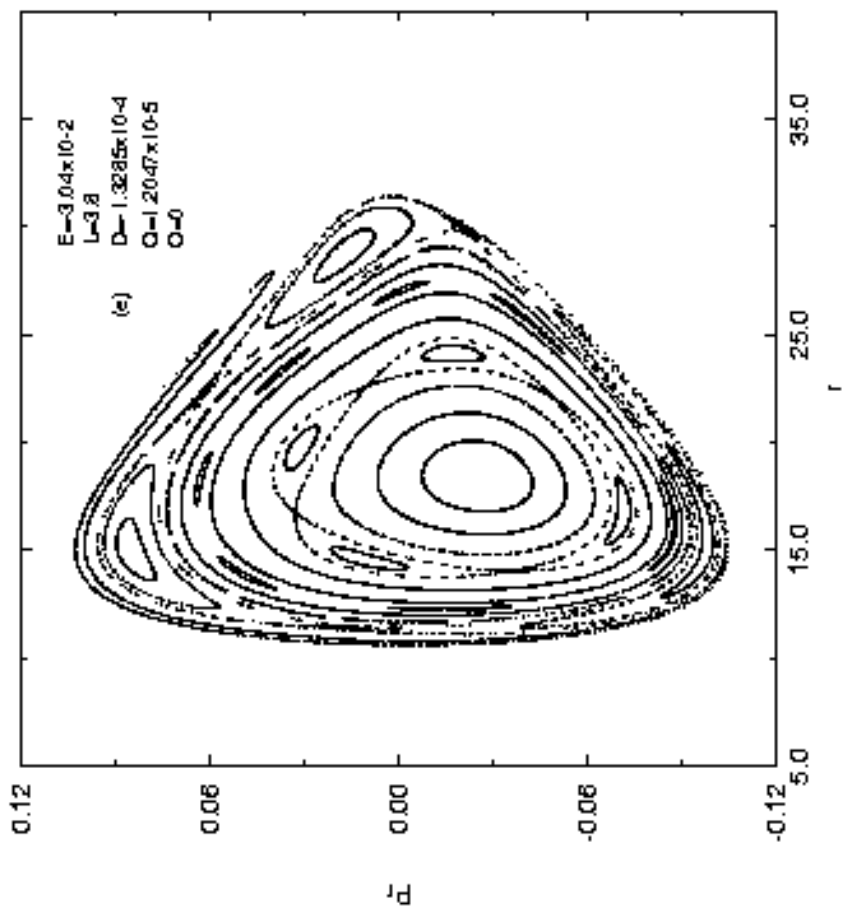
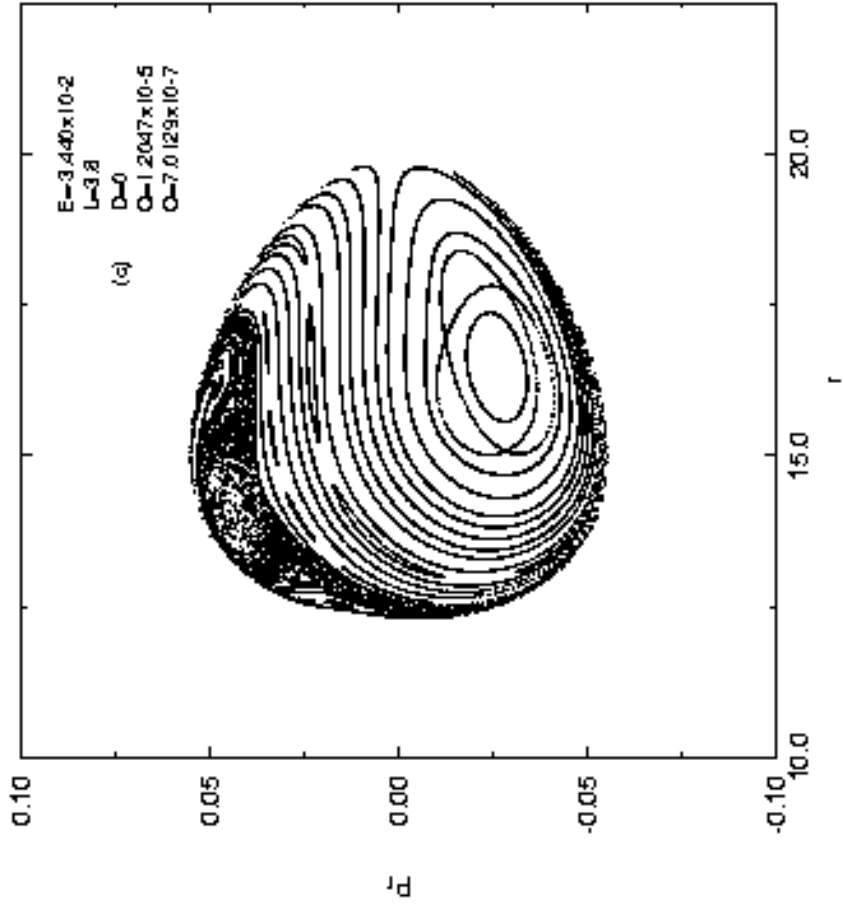
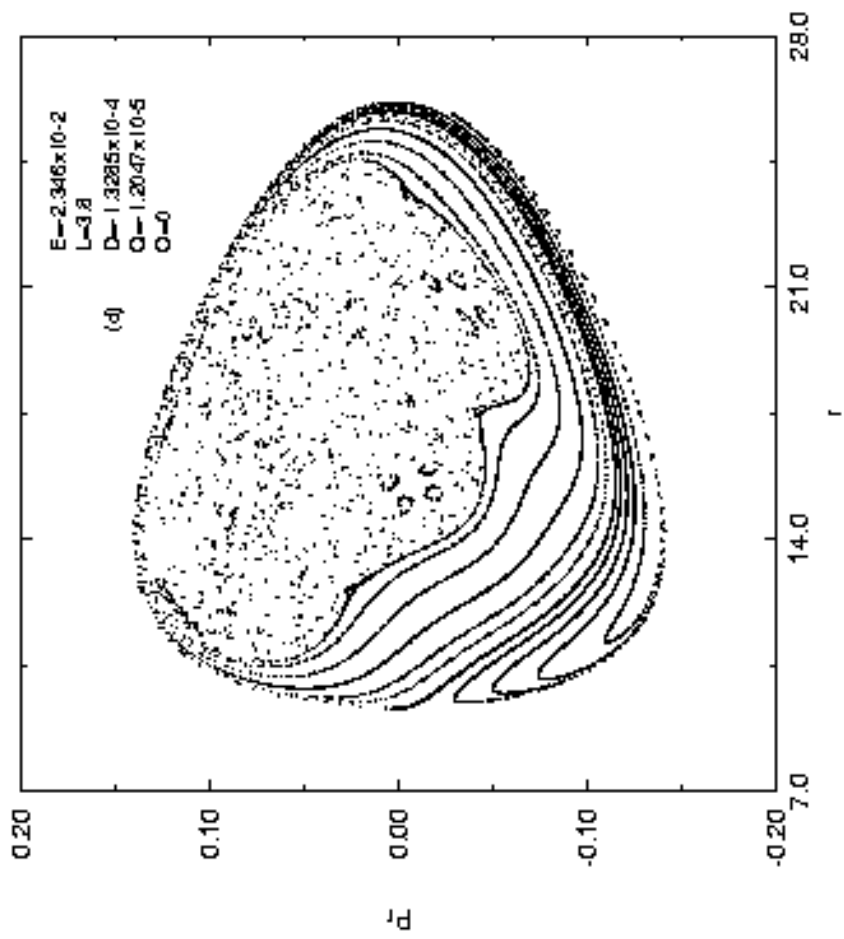
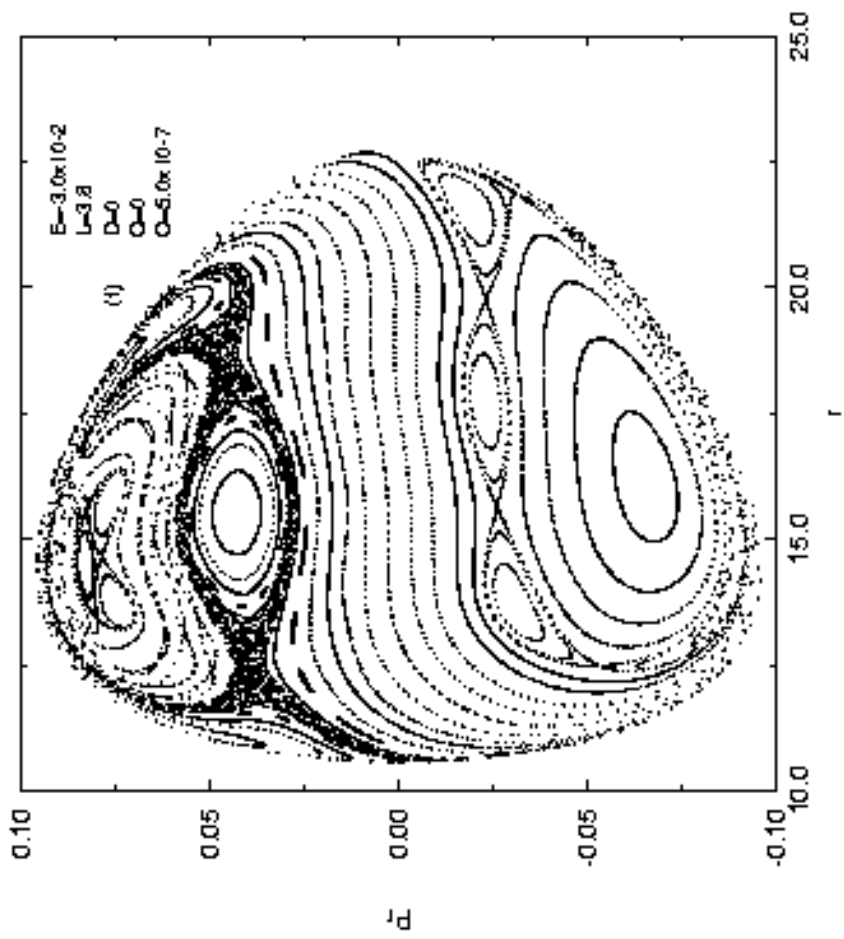
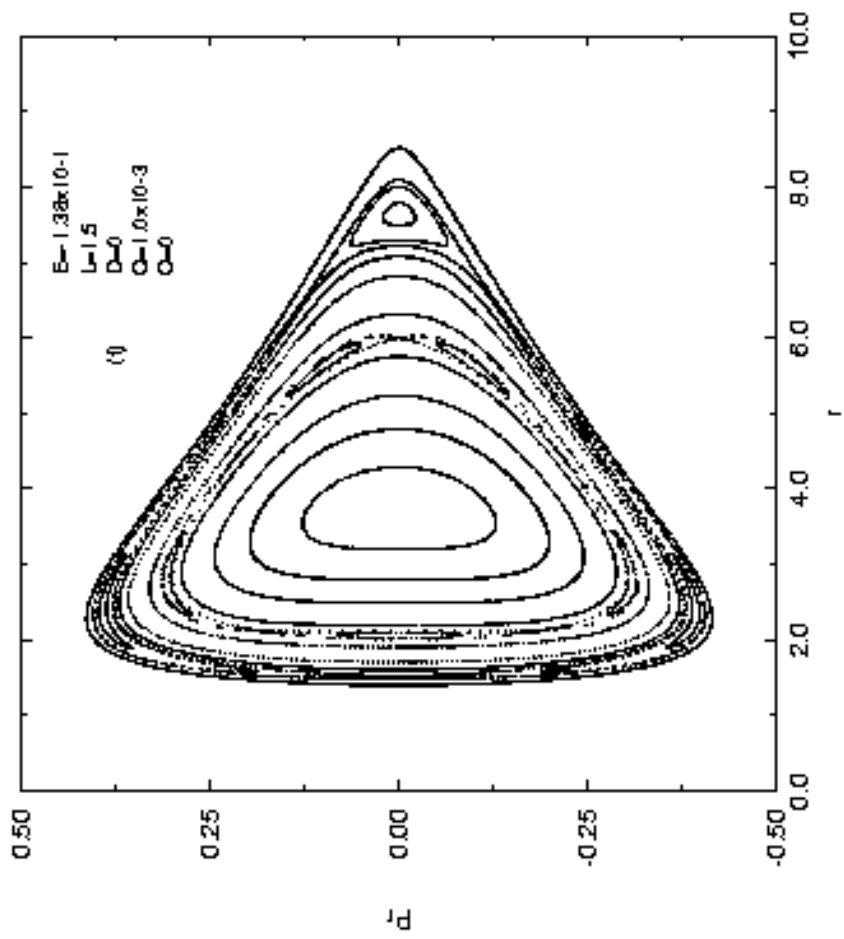
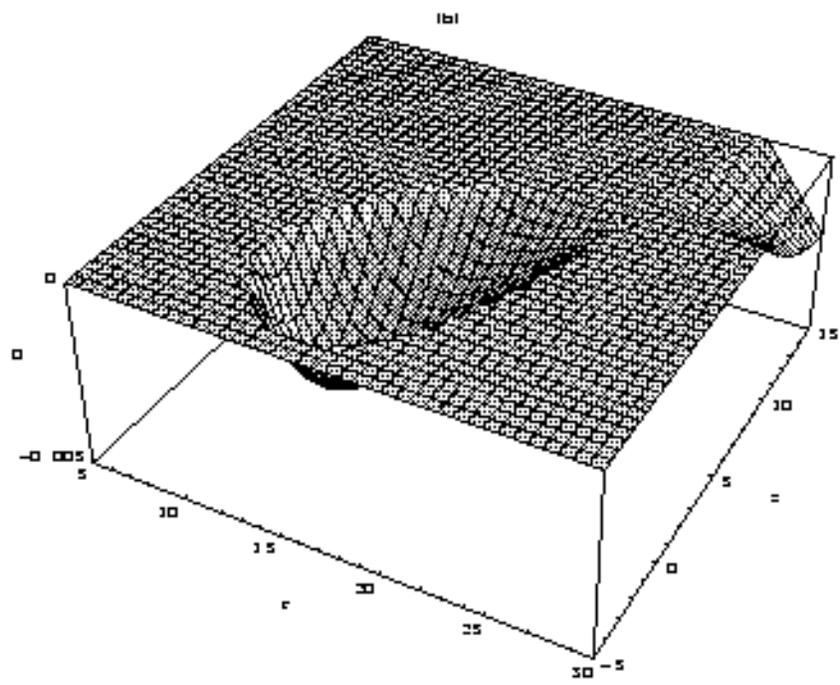
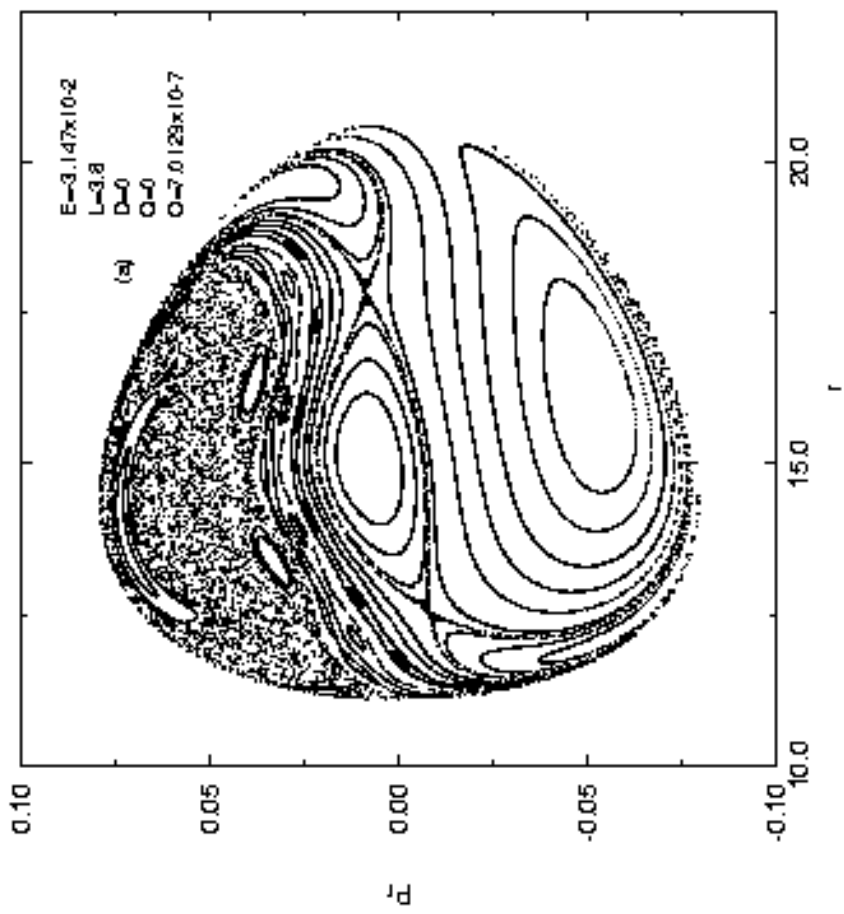
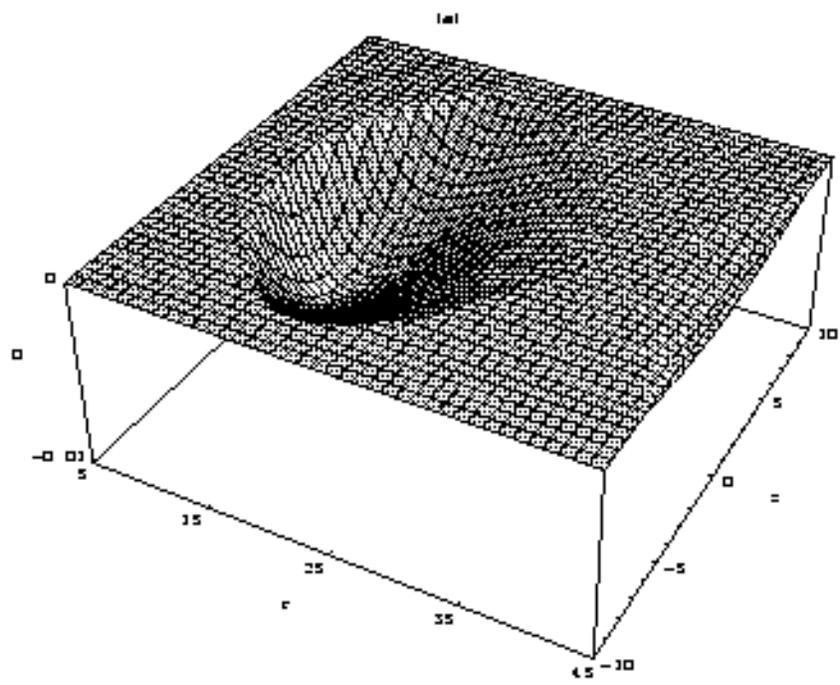
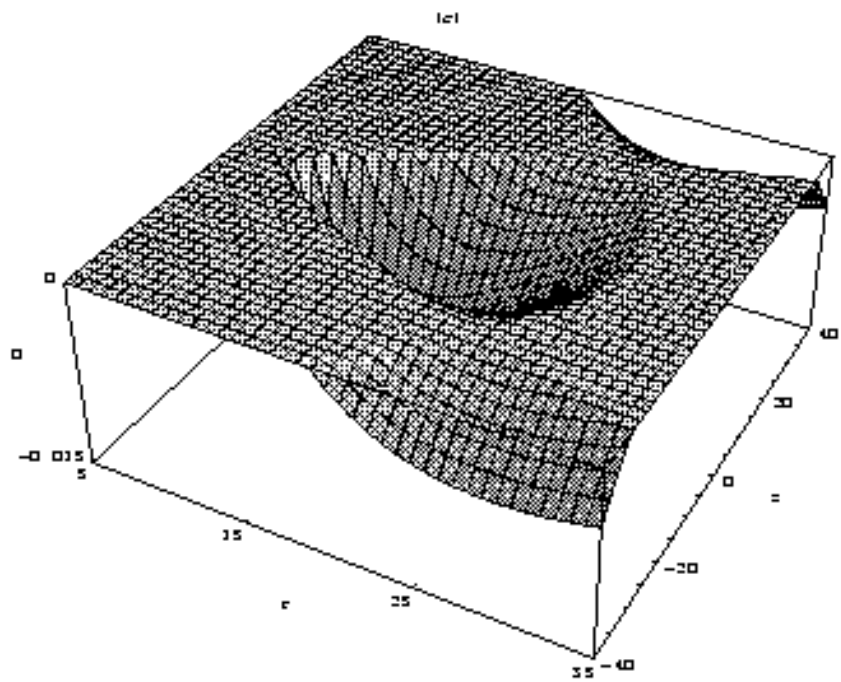
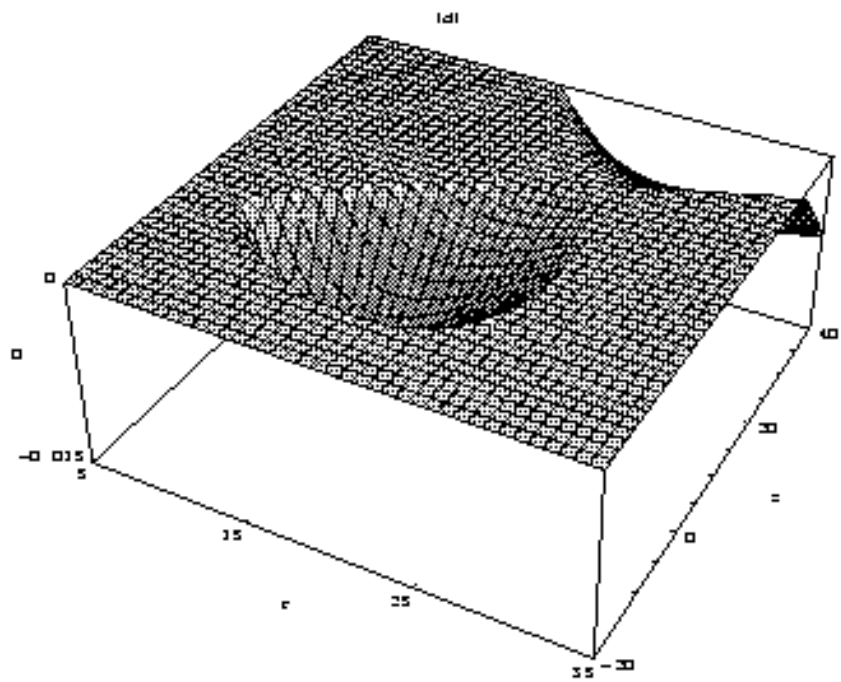
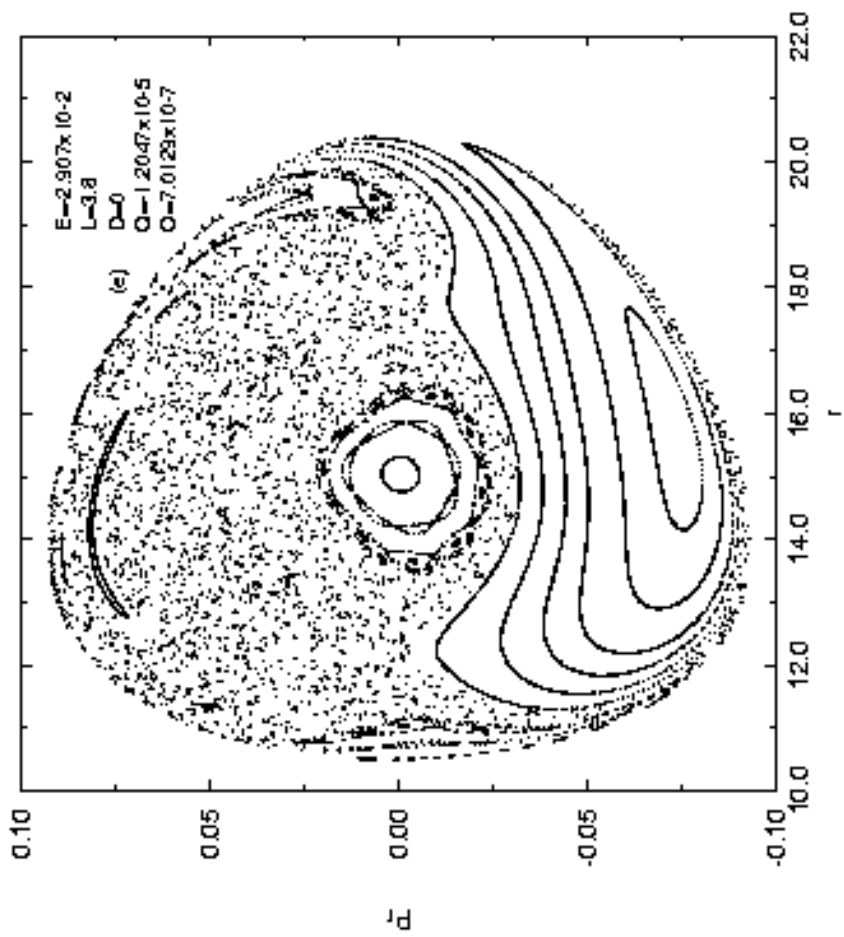
\end{document}